\documentclass[preprint, 11pt]{article}

\usepackage[T1]{fontenc}
\usepackage[utf8]{inputenc}
\usepackage[hmargin=0.9in, vmargin=1.25in]{geometry}
\usepackage[babel]{csquotes}
\usepackage{subfig}
\usepackage{graphicx}
\usepackage[colorlinks=true,citecolor=blue,urlcolor=blue]{hyperref}
\usepackage{caption}
\usepackage{amsmath}          
\usepackage{amsthm}           
\usepackage{amssymb}          
\usepackage{bm, amsfonts}     
\usepackage{xcolor}
\usepackage{fixmath}
\usepackage{tikz}
\usepackage{bm}
\usepackage{makecell}
\usepackage{mathdots}
\usepackage{algpseudocode}
\usepackage{algorithm}

\newcommand{\out}{\textup{out}}
\newcommand{\jet}{{\textup{jet}}}
\newcommand{\W}{{\mathcal W}}
\newcommand{\A}{{\mathcal A}}
\newcommand{\apdq}{\A^+ \!\!{\Delta} Q}
\newcommand{\amdq}{\A^- \!\!{\Delta} Q}
\newcommand{\Ft}{\tilde{F}}
\newcommand{\imh}{{i-1/2}}
\newcommand{\iph}{{i+1/2}}
\newcommand{\bff}{{f}}
\newcommand{\bfr}{{r}}
\newcommand{\bfF}{{F}}
\newcommand{\bfu}{{u}}
\newcommand{\entvar}{\eta'}
\newcommand{\bfq}{{Q}}
\newcommand{\bfx}{{x}}
\newcommand{\bfs}{{s}}
\newcommand{\bfn}{{n}}
\newcommand{\f}{{\mathcal{F}}}
\newcommand{\sgn}{{\rm sgn}}
\newcommand{\Rus}{{\rm Rus}}
\newcommand{\Roe}{{\rm Roe}}
\newcommand{\EV}{{\rm EV}}
\newcommand{\efp}{\psi}
\newcommand{\entflux}{g}

\newtheorem*{remark*}{Remark}{}

\title{Numerical simulation and entropy dissipative cure of the
  carbuncle instability for the shallow water circular hydraulic jump}

\author{
    David I. Ketcheson
  \thanks{
    Computer, Electrical, and Mathematical Sciences \& Engineering Division,
    King Abdullah University of Science and Technology, 4700 KAUST, Thuwal
    23955, Saudi Arabia. (\{manuel.quezada, david.ketcheson\}@kaust.edu.sa).
  }
    \and
    Manuel Quezada de Luna
  \footnotemark[1]
}
\begin{document}

\maketitle

\begin{abstract}
We investigate the numerical artifact known as a carbuncle, in the solution
of the shallow water equations.  We propose a new Riemann solver that is based
on a local measure of the entropy residual and aims to avoid carbuncles while
maintaining high accuracy.
We propose a new challenging test problem for shallow water codes, 
consisting of a steady circular hydraulic
jump that can be physically unstable.  We show that numerical methods are prone to
either suppress the instability completely or form carbuncles.
We test existing cures for the carbuncle. In our experiments, only the proposed
method is able to avoid unphysical carbuncles without suppressing the
physical instability.
\end{abstract}

\section{Introduction}

\subsection{Numerical shock instabilities}

In \cite{peery1988blunt} a numerical instability was observed to
appear near the symmetry plane in the simulation of a bow shock.
This phenomenon, subsequently dubbed ``carbuncle", has been observed by many researchers
in similar numerical experiments for the Euler equations, and many remedies
have been proposed, mainly in the form of additional numerical dissipation
\cite{quirk1997contribution,pandolfi2001numerical,dumbser2004matrix,chauvat2005shock,ismail2009proposed,shen2014stability}.
Most notably, dissipative Riemann solvers like HLLE and Rusanov suppress the carbuncle instability \cite{quirk1997contribution}.
For a recent review of numerical shock instability and work to alleviate it,
we refer to \cite[Section 2.5]{simonnumerical}.

Given the similarity of structure between the Euler equations and the shallow
water equations, it is not surprising that carbuncles appear in numerical
solutions of the latter as well \cite{kemm2014note}.
The shallow water carbuncle behaves similarly to the
Euler carbuncle; for instance, it appears when the Roe solver is used, but not when
the HLLE or Rusanov solver is used, and can also be suppressed by particular modifications
of the Roe solver \cite{kemm2014note,bader2014carbuncle}.
In this work, we propose a new Riemann solver that blends those of Rusanov and Roe
based on a local measure of the entropy residual, 
as in \cite{guermond2011entropy,guermond2018second,guermond2018well}.
Using this Riemann solver (within the second-order Lax-Wendroff-LeVeque 
finite volume scheme \cite{leveque1997wave, leveque2002finite}) supresses the formation of carbuncles 
while maintaining an accuracy similar to that of Roe solver.  

The most common test problems used to investigate carbuncle formation are
that of bow shock formation or a steady, grid-aligned planar shock.
In both of these problems, the correct behavior is the formation of a stable
shock profile without carbuncles.  This is achieved by certain methods
designed specifically to avoid carbuncles, but also by typical first-order
accurate methods.  Thus these test problems are not adequate on their own to
evaluate methods for practical calculations.
Elling \cite{elling2009carbuncle} proposed instead a problem specifically
designed to feature a carbuncle as the physical solution.  This has
been used as a test problem to identify methods that impose excessive dissipation.

Herein we introduce a new and more exacting test problem that arises in a
common physical setting.  Like some
of the problems above, it includes an equilibrium solution consisting of a
steady shock.  Similar to the Elling problem, the equilibrium is unstable.
However, the correct manifestation of the instability is different from
the carbuncle.  This allows us to distinguish schemes that yield correct
behavior from both those that are too dissipative and those that generate
carbuncles.

In the present work, we provide a test problem that possesses a genuine instability
that leads to carbuncles in many numerical approximations.  This is an ideal
test for assessing numerical methods, since neither the presence of carbuncles
nor the complete absence of instability represents the correct behavior.
This test problem is the circular hydraulic jump.

\subsection{The circular hydraulic jump}

Perhaps the first reference to the observation of the circular hydraulic jump
comes from Lord Rayleigh \cite{rayleigh1914theory}, who wrote that it
``may usually be seen whenever a stream of water from a tap strikes a horizontal
surface''.  This phenomenon that is familiar in the everyday kitchen sink, is in
fact highly nonlinear and unintuitive.  Near the jet, the flow is shallow and
supercritical, while further away it is deeper and subcritical.  The transition from supercritical
to subcritical flow
occurs in a very narrow region and takes the form of a \emph{jump} or \emph{bore}
that is roughly circular if the surface is flat; we refer to it herein as a
circular hydraulic jump (CHJ).

Early experimental work on the CHJ began some time later
\cite{kurihara1946hydraulic,tani1949water,watson1964radial}.
Watson \cite{watson1964radial} derived the jump radius
implied by Rayleigh's approach and the vertical
velocity profile in the supercritical region, assuming a no-slip
boundary condition at the bottom.  He also studied the turbulent
flow case and performed experiments.
More detailed experiments revealed different qualitative classification
of jumps  \cite{ishigai1977heat,craik1981circular}.
Although later work incorporated more physical details (such as surface tension) into the models
\cite{bush2003influence}, Bohr et. al. showed that important properties of the jump
(particularly its radius) could be reasonably predicted using a simple shallow water
model \cite{bohr1993shallow}.

While the jump is roughly circular, under appropriate conditions it may deviate
from this shape and deform rapidly and chaotically in time.
Instability of the jump was observed from fairly early on \cite{craik1981circular}.
Under special circumstances with more viscous fluids, the jump instability may lead to
the formation of curious shapes such as polygons \cite{ellegaard1998creating}, but
for a low-viscosity fluid like water the behavior is generally chaotic.
The strength of the instability increases with the jet velocity and with the
depth at the outside of the jump.  For fluids with finite viscosity, the flow
can also be completely steady for sufficiently small velocities and depths.

As we will see, carbuncles can appear in the numerical solution of the shallow
water circular hydraulic jump.  This is natural, since the solution involves a
standing shock wave.  Dealing with the carbuncle in this context is particularly
interesting and challenging, since this standing shock should (at least in an
appropriate flow regime) be unstable, and some research has suggested that the
carbuncle is the manifestation of a true physical
instability \cite{moschetta2001carbuncle,elling2009carbuncle}.

In this work, we describe and study the circular hydraulic jump as a new
test problem for shallow water discretizations.
We solve this test problem using a novel Riemann solver that supresses the 
formation of carbuncles without dissipating important features of the solution. 

\subsection{Outline}
In Section \ref{sec2}, we review some existing numerical methods for
the shallow water equations, focusing on certain Riemann solvers.
In Section \ref{sec:blended_rs}, we propose a new Riemann solver that blends
those of Roe and Rusanov in order to avoid carbuncles without being
excessively dissipative.  In Section \ref{sec:num}, we use Clawpack to
compare the performance of the new Riemann solver to existing
solvers on several standard shallow water test problems.  In Section \ref{sec:chj}, we
study the circular hydraulic jump using the newly proposed solver.
We find that although some
existing methods behave acceptably on previous test problems, they
are not capable of providing accurate solutions for the circular hydraulic jump
across the range of flow regimes we study.
Some conclusions and future directions are discussed in Section \ref{sec:conclusion}.

\section{Numerical methods for the shallow water equations} \label{sec2}
We consider the shallow water model in two horizontal dimensions:
\begin{subequations} \label{eq:sw}
\begin{align}
    h_t + (hu)_x + (hv)_y & = 0 \\
    (hu)_t + \left(hu^2 + \frac{1}{2}gh^2\right)_x + (huv)_y & = 0 \\
    (hv)_t + (huv)_x + \left(hv^2 + \frac{1}{2}gh^2\right)_y & = 0.
\end{align}
\end{subequations}
Here $h, u,$ and $v$ are respectively the depth and the $x$- and $y$-components of
velocity, which are functions of the two spatial coordinates $(x,y)$ as well as time $t$.
The gravitational force is proportional to $g$.
The system \eqref{eq:sw} can be written in vector form as
\begin{align}
    q_t + \nabla \cdot f(q) & = 0,
\end{align}
where $q=[h, hu, hv]^T$ and the flux function $f$ is defined in accordance
with \eqref{eq:sw}.

In this work we study the behavior of certain shock-capturing finite volume
methods based on the use of Riemann solvers.  For simplicity, we discuss these
solvers in the context of a 1-dimensional problem and mesh.  
In the numerical experiments in \S\ref{sec:num} we use second-order Strang
splitting \cite{strang1968construction} to extend the method to multiple dimensions.

\subsection{Wave propagation methods}\label{sec:waveprop}
Let $Q_i$ represent the average value of the set of
conserved quantities over cell $i$, which extends from $x_\imh$ to $x_\iph$.
At each time step and at each interface $x_\imh$, we approximately solve the
Riemann problem with initial states $(Q^n_{i-1},Q^n_i)$.  The approximate
solution consists of a set of traveling discontinuities or waves $\W^p_\imh$
and corresponding speeds $s^p_\imh$.
We use the wave propagation framework developed by LeVeque \cite{leveque1997wave, leveque2002finite}
and employed in the Clawpack software \cite{clawpack,pyclaw-sisc}, which implements the 
Lax-Wendroff-LeVeque (LWL) scheme
\begin{align}\label{second-order_via_fluct}
  Q_i^{n+1} = Q_i^n-\frac{\Delta t}{\Delta x}\left[\apdq_\imh+\amdq_\iph\right]
  -\frac{\Delta t}{\Delta x}\left[\tilde{F}_{i+1/2}-\tilde{F}_{i-1/2}\right],
\end{align}
Upon defining $z^+:=\frac{1}{2}(z+|z|)$ and $z^-:=\frac{1}{2}(z-|z|)$,
the fluctuations are given by
\begin{align}\label{fluct}
  \apdq_\imh := \sum_p\left(s_{i-1/2}^p\right)^+\W_{i-1/2}^p, \qquad
  \amdq_\iph := \sum_p\left(s_{i+1/2}^p\right)^-\W_{i+1/2}^p, 
\end{align}
and represent the effect (to first
order accuracy) of waves coming from the solution of the Riemann problem at
each of the neighboring interfaces.  Meanwhile, $\Ft_{i\pm 1/2}$ are correction
fluxes that make the method second-order accurate:
\begin{align}\label{correction-fluxes}
  \tilde{F}_{i\pm 1/2}=\frac{1}{2}\sum_p|s_{i\pm 1/2}^p|\left(1-\frac{\Delta t}{\Delta x}|s_{i\pm 1/2}^p|\right)\tilde\W_{i\pm 1/2}^p.
\end{align}
Here $\tilde{\W}_{i\pm 1/2}^p$ is a limited version of $\W_{i\pm 1/2}^p$, usually
based on a total variation diminishing limiter function.

We note that, for the conservative Riemann solvers we use herein, the LWL scheme
\eqref{second-order_via_fluct} can be written equivalently in flux-differencing
form:
\begin{align}\label{flux-differencing-form}
  Q_i^{n+1}=Q_i^n-\frac{\Delta t}{\Delta x}\left[\bfF_{i+1/2}-\bfF_{i-1/2}\right]
 -\frac{\Delta t}{\Delta x}\left[\tilde{F}_{i+1/2}-\tilde{F}_{i-1/2}\right],
\end{align}
with appropriately chosen first order fluxes $\bfF_{i+1/2}$ and correction
fluxes $\tilde{F}_{i-1/2}$.
We will sometimes work with
the first-order method obtained by setting the correction fluxes to zero:
\begin{align}\label{first-order_via_fluct}
  Q_i^{n+1} = Q_i^n-\frac{\Delta t}{\Delta x}\left[\apdq_\imh+\amdq_\iph\right]
 = Q_i^n -\frac{\Delta t}{\Delta x}\left[\bfF_{i+1/2}-\bfF_{i-1/2}\right].
\end{align}

Next we briefly review two commonly-used Riemann solvers: those of Rusanov and
Roe.  We refer to \cite{ketcheson2020riemann} and references therein for
a detailed description of these two Riemann solvers in the context of
the shallow water equations.
Neither of these solvers deals with the carbuncle instability in a fully
satisfactory way.  Rusanov's solver suppresses the carbuncle but 
(unless the mesh is highly refined)
is known to
also suppress related real instabilities, while Roe's solver exhibits carbuncles.
Later we will combine these two solvers in order to better deal with shock instability.
Both of these solvers satisfy the conservation property
\begin{align} \label{rs_conservation}
    \amdq_\imh + \apdq_\imh = \bff(\bfq_i) - \bff(\bfq_{i-1}).
\end{align}

\subsection{Rusanov's Riemann solver}\label{sec:rusanov}
Rusanov's solver approximates the Riemann solution with a single wave traveling in each
direction.  Both waves are assumed to travel with speed $\lambda^{\max}_\imh$, which is an
upper bound on the wave speeds appearing in the true solution of the Riemann problem.
In this work, we take $\lambda_{i-1/2}^{\max}$ to be the upper bound
from \cite[Prop. 3.7]{azerad2017well}.
The waves are then given by
\begin{align}\label{rusanov_waves}
  \W_{i-1/2}^{1,\Rus}=\bfq_{i-1/2}-\bfq_{i-1}, \qquad
  \W_{i-1/2}^{2,\Rus}=\bfq_{i}-\bfq_{i-1/2},
\end{align}
where the intermediate state $\bfu_{i-1/2}$ is determined by imposing conservation:
\begin{align}\label{rus_cons}
  -\lambda_{i-1/2}^{\max}\W_{i-1/2}^{1,\Rus}+\lambda_{i-1/2}^{\max}\W_{i-1/2}^{2,\Rus} = \bff(\bfq_{i})-\bff(\bfq_{i-1}).
\end{align}
The fluctuations are given by \eqref{fluct} with $s_{i-1/2}^1=-\lambda_{i-1/2}^{\max}$
and $s_{i-1/2}^2=\lambda_{i-1/2}^{\max}$, and the first- and second-order methods based
on Rusanov's solver are given by \eqref{first-order_via_fluct} and \eqref{second-order_via_fluct}, respectively.

\subsection{Roe's Riemann solver} \label{sec:roe}
The Roe Riemann solver is based on evaluating the flux Jacobian using Roe's average
\begin{align}\label{roe_average}
  \bar h_{i-1/2}=\frac{1}{2}(h_{i-1}+h_i), \qquad
  \hat u_{i-1/2}=\frac{\sqrt{h_{i-1}}u_{i-1}+\sqrt{h_i}u_i}{\sqrt{h_{i-1}}+\sqrt{h_i}}, \qquad
  \hat v_{i-1/2}=\frac{\sqrt{h_{i-1}}v_{i-1}+\sqrt{h_i}v_i}{\sqrt{h_{i-1}}+\sqrt{h_i}}.
\end{align}
The waves and speeds are given by the eigenvectors and eigenvalues of the approximate
flux Jacobian obtained using these averages, resulting in
\begin{align*}
  \hat\lambda_{i-1/2}^1=\hat u_{i-1/2}-\sqrt{g\bar h_{i-1/2}}, \qquad
  \hat\lambda_{i-1/2}^2=\hat u_{i-1/2}, \qquad
  \hat\lambda_{i-1/2}^3=\hat u_{i-1/2}+\sqrt{g\bar h_{i-1/2}}. \\
\end{align*}
and $\W_{i-1/2}^{p,\Roe}=\alpha_{i-1/2}^p\hat r_{i-1/2}^p$, with
\begin{align*}
  \hat \bfr^1_{i-1/2} =
  \begin{bmatrix}
    1 \\
    \hat u_{i-1/2}-\sqrt{g\bar h_{i-1/2}}\\
    \hat v_{i-1/2}
  \end{bmatrix},
  \qquad
  \hat \bfr^2_{i-1/2} =
  \begin{bmatrix}
    0\\
    0\\
    1
  \end{bmatrix},
  \qquad
  \hat \bfr^3_{i-1/2} =
  \begin{bmatrix}
    1 \\
    \hat u_{i-1/2}+\sqrt{g\bar h_{i-1/2}}\\
    \hat v_{i-1/2}
  \end{bmatrix}
\end{align*}
and the factors $\alpha^p_\imh$ obtained by solving
\begin{align}\label{system_for_alphas}
  \sum_p\W_{i-1/2}^{p,\Roe} = \left[\hat \bfr^1_{i-1/2} ~\hat \bfr^2_{i-1/2} ~\hat \bfr^3_{i-1/2}\right]
  \begin{bmatrix}
    \alpha^1_{i-1/2} \\
    \alpha^2_{i-1/2} \\
    \alpha^3_{i-1/2}
  \end{bmatrix}
  =\Delta Q_{i-1/2}:=\bfq_i-\bfq_{i-1}.
\end{align}

\subsection{Kemm's Riemann solver} \label{sec:kemm}
Several specialized Riemann solvers have been proposed in order to deal
with carbuncles.  Among them, only that proposed by Kemm has been implemented
and tested on the shallow water equations \cite{kemm2014note,bader2014carbuncle}.
The idea behind that solver, known as the HLLEMCC solver, is to combine
the HLLE and Roe solvers in such a way that the resulting method behaves
like Roe's away from potential carbuncle instabilities, and like HLLE
where the potential for such instabilities arises.   This is done using
an indicator function based on the local residual of the Rankine-Hugoniot
condition for the shear wave.  The method requires some parameters that
may need to be tuned for a specific problem.  We will consider this method
in the numerical tests below.

\section{An entropy-based blending of Rusanov and Roe}\label{sec:blended_rs}

As we discussed in the introduction, finite volume methods that use a
Roe Riemann solver are prone to exhibit the carbuncle instability.
In contrast, methods that use Rusanov's Riemann solver exhibit no carbuncles. 
However, such methods tend to dissipate other important physical features of
the solution, unless the grid is highly refined.
In this section we propose a Riemann solver
that is carbuncle-free but avoids dissipating important features of the solution.
To do so we first use a combination of Rusanov's and Roe's solvers, switching
between them based on a local measure of the entropy residual.
Previous works have also proposed blending Riemann solvers with different
amounts of dissipation in the context of the Euler equations (see
\cite{nishikawa2008very,wang2016developing,jaisankar2007diffusion,ohwada2018simple,deng2019new,ray2013entropy}),
the shallow water equations (see \cite{bader2014carbuncle,kemm2014note}),
and even the Navier-Stokes equations (see \cite{nishikawa2008very,ohwada2018simple}).
Our approach differs from those just cited in that it is based on the local entropy;
for a related approach in the context of the Euler equations, see
\cite{ismail2009affordable,ismail2009proposed}.

Since Roe's solver is not entropy stable, the blended solver is also
not guaranteed to be entropy stable.  We therefore include an additional term
that is chosen to give entropy stability, at least if the second-order correction
fluxes are neglected.  The complete proposed scheme can be written in
flux-differencing form \eqref{flux-differencing-form} using
the first-order fluxes
\begin{align} \label{blended-flux-1}
  \bfF_{i+1/2} = \frac{\bff(\bfq_{i+1})+\bff(\bfq_i)}{2}
  - \frac{1}{2} \left( \theta_{i+1/2}\lambda_{i+1/2}^{\max}\mathbb{I} + (1-\theta_{i+1/2})|\hat A_{i+1/2}| +\lambda_{i+1/2}^{\min}\mathbb{I}\right)(\bfq_{i+1}-\bfq_{i})
\end{align}
and correction fluxes \eqref{correction-fluxes} based on the Roe waves with
modified speeds
\begin{align} \label{s_p}
    s_{i+1/2}^p=\sgn(\hat\lambda_{i+1/2}^p)\lambda_{i+1/2}^p
\end{align}
where
\begin{align}\label{lambda_p}
  \lambda_{i+1/2}^p := \theta_{i+1/2}\lambda_{i+1/2}^{\max} + (1-\theta_{i+1/2})|\hat \lambda_{i+1/2}^p| + \lambda_\iph^{\min}.
\end{align}
Here $\mathbb{I}\in\mathbb{R}^{3\times 3}$ is the identity matrix,
$\lambda_{i+1/2}^{\max}$ is the
upper bound on the wave speed used in Rusanov's Riemann solver (see \S \ref{sec:rusanov}),
and $\hat A_{i+1/2}$ is Roe's averaged flux Jacobian (see \S \ref{sec:roe}).

In the following sections we describe the properties of this solver and
explain how $\theta_\iph$ and $\lambda^{\min}_\iph$ are chosen.

\subsection{The entropy residual}\label{sec:entres}
Let $\entvar(q)$ and $\entflux(q)$ denote the entropy variable and entropy flux.
Recall that $q=[h,hu,hv]^T$. We use the total energy as entropy; that is, 
\begin{align*}
  \eta(q)=\frac{1}{2} gh^2 + \frac{(hu)^2+(hv)^2}{2h}, \qquad
  \entflux(q)=
    \frac{\eta(q)}{h}
    \begin{bmatrix} hu \\ hv \end{bmatrix}.
\end{align*}
Based on \cite{guermond2011entropy}, we consider
\begin{align}\label{ent_residual}
  \int_{S_i} \entvar(q) \cdot \left[ \frac{\partial q}{\partial t} + \nabla\cdot \bff(q)\right]d\bfx
  =\int_{S_i} \left[\frac{\partial \eta(q)}{\partial t} + \entvar(q) \cdot \nabla\cdot \bff(q)\right] d\bfx
\end{align}
as a measurement of the entropy production in cell $i$ (here $S_i=[x_\imh,x_\iph]$).
To avoid the need to compute the time derivative of the entropy, we follow
\cite{guermond2018second, guermond2018well} and use
$\int \frac{\partial\eta(q)}{\partial t}d\bfx=-\int \nabla\cdot\entflux(q) d\bfx$,
which holds in smooth regions. Then we define
\begin{align}\label{Ri}
  \theta_i := \frac{R_i}{D_i},
\end{align}
with
\begin{align*}
  R_i=
  \left|\int_{S_i} \left[\entvar(q) \cdot \nabla\cdot \bff(q) - \nabla\cdot\entflux(q) \right] d\bfx\right|,
\end{align*}
and $D_i$ being an upper bound on $R_i$ so that $0\leq \theta_i\leq 1$.
Note that $R_i\approx 0$ if $q$ is smooth in $S_i$.
In our implementation, we use the approximation
\begin{align}
  R_i 
  &\approx 
  \left|\entvar(\bfq_i)\cdot \int_{S_i}\nabla\cdot \bff(q)d\bfx
  -\int_{S_i}\nabla\cdot\entflux(q) d\bfx\right| 
  = 
  \left|\entvar(\bfq_i)\cdot \int_{\partial S_i}\bfn_i(s) \cdot \bff(q) d\bfs
  -\int_{\partial S_i}\bfn_i(s)\cdot\entflux(q) d\bfs\right|,
\end{align}
where $\bfn_i$ denotes the unit vector normal to $\partial S_i$ pointing outward from cell $i$. 
To compute the boundary integrals, we approximate $q$ on each cell edge $\f_{ij}$ by the average of the two
neighboring cell averages $\bar Q_{ij} = (Q_i+Q_j)/2$, resulting in
\begin{subequations}\label{Ri_and_Di}
\begin{align}
  R_i = 
  \left|\sum_{\f_{ij}\in\partial S_i} |\f_{ij}|
  \left[\entvar(\bfq_i)\cdot \left(\bfn_{ij} \cdot \bff\left(\bar Q_{ij}\right)\right)-\bfn_{ij} \cdot\entflux \left(\bar Q_{ij}\right)\right] \right|,
\end{align}
where $|\f_{ij}|$ is the length of $\f_{ij}$ and 
$\bfn_{ij}$ is the unit vector normal to $\f_{ij}$ pointing outward from cell $i$.
The normalization factor is similarly computed as
\begin{align}
  D_i 
  = 
  \sum_{k=1}^{d+1}\left|\entvar_k\left(\bfq_i\right)\right|
  \left|\sum_{\f_{ij}\in\partial S_i}|\f_{ij}|\left(\bfn_{ij}\cdot\bff\left(\bar Q_{ij}\right)\right)_k\right|
  +\left|\sum_{\f_i\in \partial S_i} |\f_{ij}| \bfn_{ij}\cdot\entflux\left(\bar Q_{ij}\right) \right|,
\end{align}
\end{subequations}
where $(z)_k$ denotes the $k$-th component of $z\in\mathbb{R}^{d+1}$, 
and $d$ is the number of physical dimensions.
In \eqref{blended-flux-1} we need values of $\theta$ at the cell interfaces, for
which we use
$$
    \theta_\iph = \max(\theta_i, \theta_{i+1}).
$$
We expect that $\theta_\iph\approx 0 $ in smooth regions,
while $\theta_\iph\approx 1$ near shocks.
The value of $\theta_\iph$ controls whether the first-order flux \eqref{blended-flux-1}
behaves more like that of Rusanov or Roe.  Specifically, if
$\lambda_\iph^{\min}=0$ and $\theta_\iph=1$, the flux is equivalent to that of
Rusanov, while if $\lambda_\iph^{\min}=0$ and $\theta_\iph=0$, it is equivalent
to that of Roe.  

In fact, with the choice \eqref{s_p}, the correction fluxes also match
those of Rusanov or Roe in the appropriate limit, as shown in the next section.

\subsection{The correction fluxes}
In this section we explain the choice of wave speeds \eqref{s_p}.
For the moment, we consider \eqref{blended-flux-1}-\eqref{lambda_p} without
entropy stabilization; i.e., we set $\lambda^{\min}_\iph=0$
for now.
We use $\W^{p,\Rus}$ to denote the waves in the Rusanov
solver and $\W^{p,\Roe}$ to denote the waves in the Roe solver.
The first-order method given by \eqref{first-order_via_fluct} with the numerical
flux \eqref{blended-flux-1} can also be written in
terms of fluctuations:
\begin{subequations}\label{ev_fluctuations}
\begin{align}
  \apdq_\imh & = \frac{1}{2}\sum_p \left(\hat\lambda_{i-1/2}^p + \lambda_{i-1/2}^p\right)\W_{i-1/2}^{p,\Roe}, \\
  \amdq_\iph & = \frac{1}{2}\sum_p \left(\hat\lambda_{i+1/2}^p - \lambda_{i+1/2}^p\right)\W_{i+1/2}^{p,\Roe}.
\end{align}
\end{subequations}
To implement the correction fluxes required for the second-order scheme \eqref{second-order_via_fluct},
we must also define a set of waves and corresponding speeds.
Using only the waves from the Roe solver, it is in general not possible
to choose speeds that yield the fluctuations \eqref{ev_fluctuations},
and we instead use \eqref{s_p}.
Using this in \eqref{correction-fluxes} and \eqref{ev_fluctuations} in 
\eqref{second-order_via_fluct}, we obtain (in the absence of limiting)
the second-order scheme
\begin{align}\label{LWL}
  \bfq_i^{n+1}=\bfq_i^n
  &-\frac{\Delta t}{2 \Delta x}
  \sum_p
  \left[\left(\hat\lambda_{i+1/2}^p -\frac{\Delta t}{\Delta x}(\lambda_{i+1/2}^p)^2 \right)\W_{i+1/2}^{p,\Roe}
  +
  \left(\hat\lambda_{i-1/2}^p +\frac{\Delta t}{\Delta x}(\lambda_{i-1/2}^p)^2 \right)\W_{i-1/2}^{p,\Roe}\right],
\end{align}
It is clear again that if $\theta_{i\pm 1/2}=0$ we recover the standard
second-order Lax-Wendroff method based on Roe's Riemann solver.
We now show that if $\theta_{i\pm 1/2}=1$, we recover the Lax-Wendroff method
based on Rusanov's Riemann solver.

To see this, first consider \eqref{rus_cons} and rewrite the right-going fluctuation as follows:
\begin{align}\label{rus_rs_aux}
  \A^{+,\Rus}\Delta Q_{i-1/2}:=\lambda_{i-1/2}^{\max}\W_{i-1/2}^{2,\Rus}
  = \bff(\bfq_{i})-\bff(\bfq_{i-1}) + \lambda_{i-1/2}^{\max}\left(\W_{i-1/2}^{1,\Rus}+\W_{i-1/2}^{2,\Rus}\right)
  -\lambda_{i-1/2}^{\max}\W_{i-1/2}^{2,\Rus}.
\end{align}
From \eqref{rusanov_waves} and \eqref{system_for_alphas} we get
$\W^{1,\Rus}_{i-1/2}+\W^{2,\Rus}_{i-1/2}=\bfq_i-\bfq_{i-1}=\sum_p\W_{i-1/2}^{p,\Roe}$.
Using \eqref{rs_conservation} and \eqref{ev_fluctuations}, \eqref{rus_rs_aux} becomes
\begin{align*}
  \lambda_{i-1/2}^{\max}\W_{i-1/2}^{2,\Rus}
  = \sum_p \left(\hat \lambda_{i-1/2}^p + \lambda_{i-1/2}^{\max}\right)\W_{i-1/2}^{p,\Roe}
  -\lambda_{i-1/2}^{\max}\W_{i-1/2}^{2,\Rus},
\end{align*}
which implies that
\begin{subequations}\label{fluct_rus}
\begin{align} 
  \A^{+,\Rus}\Delta Q_{i-1/2}=\frac{1}{2}\sum_p\left(\hat \lambda_{i-1/2}^p + \lambda_{i-1/2}^{\max}\right)\W_{i-1/2}^{p,\Roe},
\end{align}
and similarly,
\begin{align}
  \A^{-,\Rus}\Delta Q_{i+1/2}=\frac{1}{2}\sum_p\left(\hat \lambda_{i+1/2}^p - \lambda_{i+1/2}^{\max}\right)\W_{i+1/2}^{p,\Roe}.
\end{align}
\end{subequations}
To get the second-order Lax-Wendroff method based on Rusanov's Riemann solver,
plug the fluctuations \eqref{fluct_rus} into \eqref{second-order_via_fluct} where 
$\tilde F_{i\pm 1/2}$ is given by \eqref{correction-fluxes} with $|s^p_{i\pm 1/2}|=\lambda_{i\pm 1/2}^{\max}$ and $\tilde\W^p_{i\pm 1/2}=\W_{i\pm 1/2}^{p,\Roe}$.
By doing this, we get \eqref{LWL} (since $\theta_{i\pm 1/2}=1 \implies \lambda_{i\pm 1/2}^p=\lambda_{i\pm 1/2}^{\max}$).

\subsection{Entropy stabilization}\label{sec:entropy_stable}
In the previous section we neglected the term $\lambda^{\min}_\iph$ in
\eqref{blended-flux-1}.  In this section, we follow \cite{kuzmin2020algebraic}
and show how this term is computed, in order to guarantee local entropy stability
of the scheme given by using \eqref{blended-flux-1} in \eqref{first-order_via_fluct}.

Let $\entvar_i=\entvar(\bfq_i)$ and $\entflux_i=\entflux(\bfq_i)$ denote the entropy variable and the
(one-dimensional) entropy flux at cell $i$, respectively.
Also let $\efp=\entvar(\bfq_i)\cdot \bff(\bfq_i)-\entflux(\bfq_i)$ be the entropy potential at cell $i$.
From \cite[\S 4]{tadmor1987numerical}, if the numerical fluxes satisfy
\begin{align}\label{es_cond}
(\entvar_{i}-\entvar_{i-1})\cdot \bfF_{i-1/2}\leq \efp_{i}-\efp_{i-1},
  \qquad
  (\entvar_{i+1}-\entvar_i)\cdot \bfF_{i+1/2}\leq \efp_{i+1}-\efp_i,
\end{align}
then we have the entropy inequality
\begin{align}\label{ent_ineq}
  \frac{d\eta(\bfq_i)}{dt}+\frac{1}{\Delta x}\left[G_{i+1/2}-G_{i-1/2}\right]\leq 0,
\end{align}
where
\begin{align*}
    G_{i+1/2} &= \frac{1}{2}\left(\entvar_i+\entvar_{i+1}\right)\bfF_{i+1/2}-\frac{1}{2}(\efp_{i}+\efp_{i+1}),
\end{align*}
is a discretization of the entropy flux.
If \eqref{ent_ineq} holds with equality, the scheme is entropy-conservative
\cite{tadmor2003entropy}.  

The approach in \cite{tadmor2003entropy}, is based on first developing an
entropy-conservative scheme and then adding entropy dissipation to enforce \eqref{ent_ineq}.
On the other hand, herein we have added dissipation (as described in Section \ref{sec:entres})
that does not guarantee \eqref{ent_ineq}.
Indeed, some linear stabilization techniques that add artificial dissipation of the conserved 
variables are known to produce entropy; see for example \cite{ern2013weighting,kuzmin2020entropy}. 
To guarantee \eqref{ent_ineq}, we add extra dissipation of the conserved variables via 
\eqref{lambda_min}. 
Doing this counteracts entropy production created by any component of the Riemann solver;
see for example \cite{kuzmin2020entropy} (in the context of $C^0$ finite elements). 

Plugging \eqref{blended-flux-1} into the condition \eqref{es_cond} yields the required value
\begin{align}\label{lambda_min}
  \lambda_{i+1/2}^{\min} 
  = \max\left\{0,\frac{\frac{1}{2}(\entvar_{i+1}-\entvar_{i})\cdot
    \left[\bff(\bfq_{i+1})+\bff(\bfq_{i})-\sum_p\lambda_{i+1/2}^{p,\EV}\W_{i+1/2}^{p,\Roe}\right]
    -(\efp_{i+1}-\efp_i)}
  {\frac{1}{2}(\entvar_{i+1}-\entvar_{i})\cdot\sum_p\W_{i+1/2}^{p,\Roe}}\right\},
\end{align}
where 
\begin{align*}
  \lambda_{i+1/2}^{p,\EV}=\theta_{i+1/2}\lambda_{i+1/2}^{\max}+(1-\theta_{i+1/2})|\hat{\lambda}_{i+1/2}^p|.
\end{align*}
With this choice, \eqref{blended-flux-1} guarantees \eqref{es_cond}, and therefore \eqref{ent_ineq}.
Note that here we have used the identities
\begin{align*}
|\hat{A}_\iph|(Q_{i+1}-Q_i) = \sum_p |\hat{\lambda}^p_\iph| \W^{p,\Roe}_\iph, \qquad
Q_{i+1}-Q_i & = \sum_p \W^{p,\Roe}_\iph.
\end{align*}

Since the blended Riemann solver described in Section \ref{sec:blended_rs} already
tends to introduce entropy dissipation,
we expect condition \eqref{es_cond} to be fulfilled most of the time even with $\lambda_{i+1/2}^{\min}=0$.
But \eqref{lambda_min} is used as a safeguard to guarantee entropy stability of the first-order method.
Extra modifications would be needed to guarantee entropy stability of the second-order LWL method
\eqref{second-order_via_fluct} and its multidimensional extension via Strang splitting.
We do not pursue such modifications in this work.

We close this section with two remarks.  First, note that 
the entropy stability condition \eqref{es_cond} can be written equivalently in terms of fluctuations:
\begin{align*}
  (\entvar_{i}-\entvar_{i-1})\cdot
  \Big(
  \underbrace{\bff(\bfq_i)-\A^+\Delta Q_{i-1/2}}_{=\bfF_{i-1/2}}\Big)\leq \efp_{i}-\efp_{i-1},
  \quad
  (\entvar_{i+1}-\entvar_i)\cdot
  \Big(\underbrace{\bff(\bfq_i)+\A^-\Delta Q_{i+1/2}}_{=\bfF_{i+1/2}}\Big)\leq \efp_{i+1}-\efp_i.
\end{align*}

Second, the additional dissipation introduced by $\lambda_{i\pm 1/2}^{\min}$ can also
serve independently as an entropy fix for Roe's solver, as we demonstrate
via a numerical experiment in \S \ref{sec:rp_dry_bed}.

\section{Numerical results}\label{sec:num}
In this section we present one- and two-dimensional numerical experiments to demonstrate the behavior of the
blended Riemann solver from \S\ref{sec:blended_rs} with the extra entropy dissipation from \S\ref{sec:entropy_stable}.
We compare the behavior of the blended solver against the standard Roe's and Rusanov's solvers.
In most of the experiments we use these Riemann solvers with the LWL method reviewed in \S\ref{sec:waveprop}.
When the exact solution is available, we report convergence results based on the $L^1$-error for the water height
\begin{align*}
  E_1=\sum_i|K_i|~|h_i-h^e(\bfx_i)|,
\end{align*}
where $|K_i|$ is the length or area of cell $i$,
$h_i$ is the cell average of the water height at cell $i$,
and $h^e(\bfx_i)$ is the exact water height evaluated at the center of cell $i$.
Since the methods under consideration are at most second order accurate, comparison
of cell averages with centered point values is a sufficient way to test their accuracy.
In all experiments we use $g=1$.

We consider a total of five problems. We start in \S \ref{sec:rp_dry_bed} with a
one-dimensional Riemann problem over a dry bed.
For this problem we use the first-order methods \eqref{first-order_via_fluct},
to avoid negative depth values. Although only Rusanov's solver is proven to guarantee
positivity, we do not get negative values for the water height with any of the first-order methods.
In \S \ref{sec:rp_wet_bed} we apply the second-order methods to a dam-break problem
with a wet bed. This problem contains strong shocks. We observe similar accuracy with the
blended solver or the Roe solver; this suggests that the extra numerical entropy
dissipation that the blended solver introduces near the shocks does not degrade the accuracy
of the underlying Roe solver. This extra dissipation, however, prevents the formation of carbuncles in other experiments,
which we demonstrate in \S \ref{sec:bow_shock} and \S\ref{sec:2D_chj}.
In \S \ref{sec:steady_outflow} we consider a one-dimensional problem with a smooth steady state solution. We observe the expected
second-order accuracy of method \eqref{second-order_via_fluct} with Roe's and the blended solvers. Using
Rusanov's Riemann solver degrades the accuracy to first-order. The overall accuracy of the blended solver
is not degraded since the extra dissipation is not applied in smooth regions.
The main focus of this work is in the formation of carbuncle instabilities in the two-dimensional CHJ. We present
an extensive set of experiments for this problem in \S \ref{sec:2D_chj}.
We consider not only Roe's, Rusanov's and the blended solvers,
but also the solver proposed in \cite{kemm2014note},
which is designed to remove the carbuncle instabilities in the shallow water equations.

\subsection{Dam break problem on a dry bed}\label{sec:rp_dry_bed}
\begin{figure}[!h]
  {\scriptsize
    \subfloat[Roe's Riemann solver.\label{fig:rp_dry_bed_roe}]{
      \includegraphics[scale=0.36]{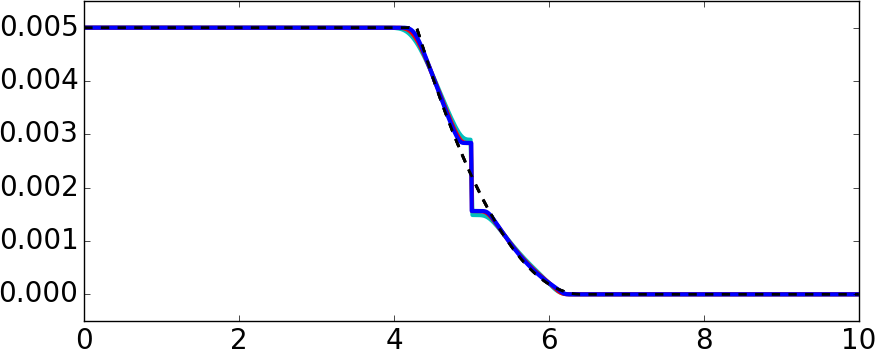}}
    ~
    \subfloat[Rusanov's Riemann solver.\label{fig:rp_dry_bed_llf}]{
        \includegraphics[scale=0.36]{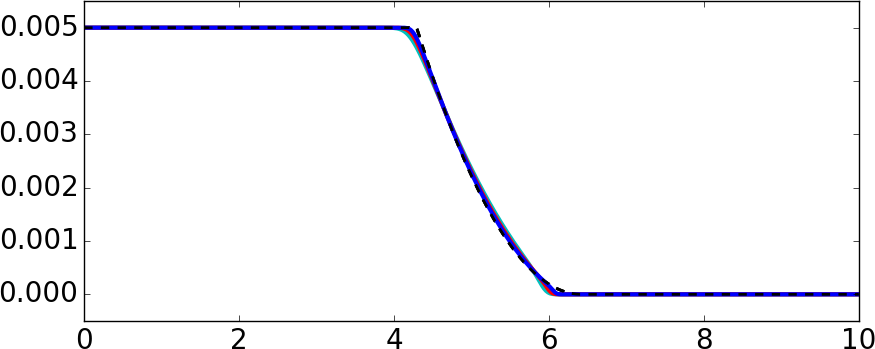}}

    \subfloat[Blended Riemann solver. \label{fig:rp_dry_bed_blended}]{
      \includegraphics[scale=0.36]{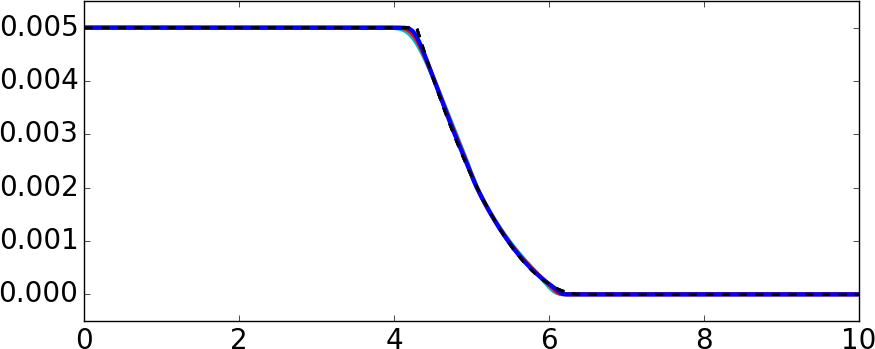}}
    ~
    \subfloat[Roe's Riemann solver with $\lambda_{i\pm 1/2}^{\min}$.\label{fig:rp_dry_bed_roe_with_lmin}]{
      \includegraphics[scale=0.36]{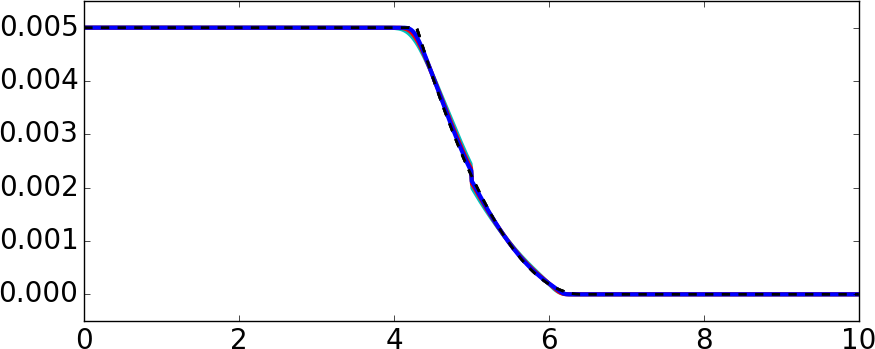}}
  }
  \caption{
    Dam break problem over a dry bed using method \eqref{first-order_via_fluct} with different
    Riemann solvers. We show the numerical solution and the exact solution (in dashed black) at $t=10$.
    We consider different refinements with
    (cyan) $\Delta x=1/400$, (red) $\Delta x=1/800$, (blue) $\Delta x=1/1600$.
    \label{fig:rp_dry_bed}}
\end{figure}

We start with the one-dimensional dam break problem on a dry bed.
This problem is a canonical example that demonstrates the `entropy glitch' of Roe's solver
at transonic rarefactions. Since the baseline Roe solver we use (see \S\ref{sec:roe}) does not contain
an entropy fix, it is important to demonstrate that the blended Riemann solver fixes the entropy glitch.
We follow the setup in \cite[\S 4.1.2]{delestre2013swashes}.
The domain is given by $x\in(0,10)$, and the initial condition is $hu(x,0)=0$ and
\begin{align*}
  h(x,0)=
  \begin{cases}
    h_l & x\leq x_0=5 \\
    h_r & x >x_0,
  \end{cases}
\end{align*}
with $h_l=5\times 10^{-3}$.  In our experiments we use $h_r=1\times
10^{-15}$ to avoid division by zero.
The exact solution, which can also be found in \cite{delestre2013swashes} and references therein, is
\begin{align*}
  h(x,t) =
  \begin{cases}
    h_l, \\
    \frac{4}{9g}\left(\sqrt{gh_l}-\frac{x-x_0}{2t}\right)^2, \\
    0,
  \end{cases}
\quad
  u(x,t) =
  \begin{cases}
    0, &\mbox{ if } x\leq x_A(t), \\
    \frac{2}{3}\left(\sqrt{gh_l}+\frac{x-x_0}{t}\right), & \mbox{ if } x_A(t) < x\leq x_B(t), \\
    0, &\mbox{ if } x_B(t) < x,
  \end{cases}
\end{align*}
with $x_A(t)=x_0 - t\sqrt{gh_l}$ and $x_B=x_0+2t\sqrt{gh_l}$.
We solve the problem up to $t=10$.
In Figures \ref{fig:rp_dry_bed_roe}-\ref{fig:rp_dry_bed_blended},
we show the solution using method \eqref{first-order_via_fluct}
with Roe's solver, Rusanov's solver and the blended solver, respectively.
The entropy glitch, which is manifested as a non-physical shock at $x=x_0$,
is present with Roe's solver. Using Rusanov's and the blended solvers
fixes the entropy glitch.
In Table \ref{table:rp_dry_bed}, we summarize the results of a convergence test.
Note that using the blended solver leads to more accurate results.

For this particular problem, using method \eqref{first-order_via_fluct} with
the blended Riemann solver leads to $\lambda_{i\pm 1/2}^{\min}=0$ for all the experiments.
We can artificially activate the entropy stabilization by imposing $\theta_{i}=0$ in \eqref{lambda_p},
which is equivalent to using Roe's solver with extra dissipation given by $\lambda_{i\pm 1/2}^{\min}$.
In Figure \ref{fig:rp_dry_bed_roe_with_lmin} we show the solution,
and in the last column of Table \ref{table:rp_dry_bed} we summarize the results of a convergence test.
The entropy glitch is still noticeable but greatly reduced compared to the
solution from Roe's method without entropy stabilization.
To remove completely the entropy glitch we could add high-order entropy dissipation following \cite{tadmor2003entropy}
and references therein.

\begin{table}[!ht]\scriptsize
  \begin{center}
    \begin{tabular}{||c||c|c||c|c||c|c||c|c||} \hline
      & \multicolumn{2}{c||}{Roe's solver}
      &\multicolumn{2}{c||}{Rusanov's solver}
      &\multicolumn{2}{c||}{Blended solver}
      &\multicolumn{2}{c||}{Roe's with $\lambda_{i\pm 1/2}^{\min}$} \\ \cline{2-9}
      $\Delta x$ & $E_1$ & rate & $E_1$ & rate & $E_1$ & rate & $E_1$ & rate \\ \hline
      1/50   & 6.21E-04 &  --  & 6.95E-04 &   -- & 5.08E-04 & --   & 5.95E-04 & --   \\
      1/100  & 4.70E-04 & 0.40 & 5.27E-04 & 0.40 & 3.39E-04 & 0.58 & 4.06E-04 & 0.55 \\
      1/200  & 3.91E-04 & 0.26 & 3.88E-04 & 0.44 & 2.30E-04 & 0.56 & 2.83E-04 & 0.52 \\
      1/400  & 3.16E-04 & 0.30 & 2.64E-04 & 0.55 & 1.49E-04 & 0.62 & 1.89E-04 & 0.58 \\
      1/800  & 2.35E-04 & 0.42 & 1.67E-04 & 0.66 & 9.24E-05 & 0.68 & 1.20E-04 & 0.65 \\
      1/1600 & 2.00E-04 & 0.23 & 1.01E-04 & 0.72 & 5.66E-05 & 0.70 & 7.42E-05 & 0.68 \\ \hline
    \end{tabular}
    \caption{Grid convergence study for the dam break problem over a dry bed
      using method \eqref{first-order_via_fluct} with different Riemann solvers.\label{table:rp_dry_bed}}
  \end{center}
\end{table}

\subsection{Dam break problem on a wet bed}\label{sec:rp_wet_bed}
\begin{figure}[!h]
  {\scriptsize
    \subfloat[Roe's Riemann solver.]{
      \includegraphics[scale=0.29]{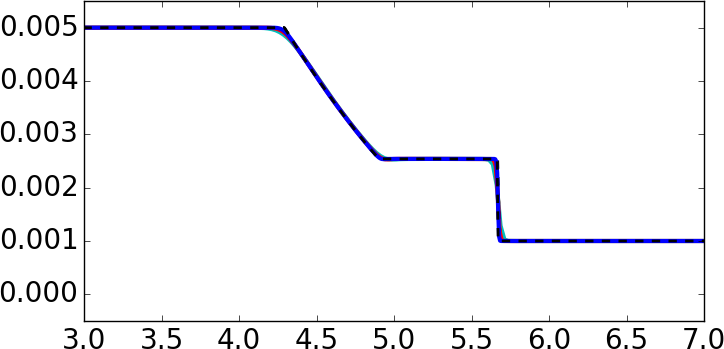}}
    \quad
    \subfloat[Rusanov's Riemann solver.]{
        \includegraphics[scale=0.29]{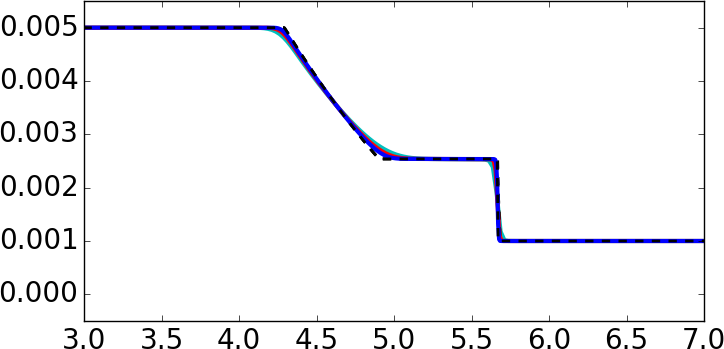}}
    \quad
    \subfloat[Blended Riemann solver.]{
      \includegraphics[scale=0.29]{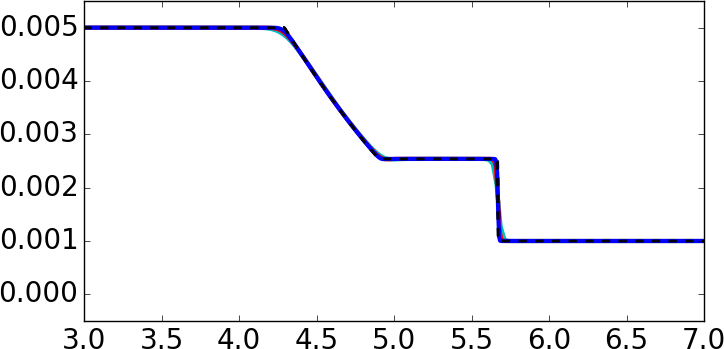}}
  }
  \caption{
    Dam break problem over a wet bed using method \eqref{second-order_via_fluct} with different
    Riemann solvers. We show the numerical solution and the exact solution (in dashed black) at $t=10$.
    We consider different refinements with
    (cyan) $\Delta x=1/400$, (red) $\Delta x=1/800$, (blue) $\Delta x=1/1600$.
    \label{fig:rp_wet_bed}}
\end{figure}

We consider now a one-dimensional dam break problem on a wet domain.
We follow the setup in \cite[\S 4.1.1]{delestre2013swashes}.
The domain is $x\in(0,10)$ and the initial condition is given by
$hu(x,0)=0$ and
\begin{align*}
  h(x,0) =
  \begin{cases}
    h_l & x \leq x_0 \\
    h_r & x > x_0
  \end{cases}
\end{align*}
with $x_0=5$, $h_l=0.005$, and $h_r=0.001$.  The exact solution, which can be found in
\cite{delestre2013swashes} and references therein, is given by
\begin{align*}
  h(x,t) =
  \begin{cases}
    h_l, \\
    \frac{4}{9g}\left(\sqrt{gh_l}-\frac{x-x_0}{2t}\right)^2, \\
    \frac{c_m^2}{g}, \\
    h_r,
  \end{cases}
\quad
  u(x,t) =
  \begin{cases}
    0, &\mbox{ if } x\leq x_A(t), \\
    \frac{2}{3}\left(\sqrt{gh_l}+\frac{x-x_0}{t}\right), & \mbox{ if } x_A(t) < x\leq x_B(t), \\
    2(\sqrt{gh_l}-c_m), & \mbox{ if } x_B(t)<x\leq x_C(t), \\
    0, &\mbox{ if } x < x_C(t),
  \end{cases}
\end{align*}
where $x_A(t)=x_0-t\sqrt{gh_l}$, $x_B(t)=x_0+t\left(2\sqrt{gh_l}-3c_m\right)$,
$x_C(t)=x_0+t\frac{2c_m^2\left(\sqrt{gh_l}-c_m\right)}{c_m^2-gh_r}$ and
$c_m$ is the solution of
$-8gh_rc_m^2\left(\sqrt{gh_l}-c_m\right)^2+\left(c_m^2-gh_r\right)^2\left(c_m^2+gh_r\right)=0$.
We solve the problem up to the final time $t=5$ using the
second-order method \eqref{second-order_via_fluct}
with Roe's, Rusanov's and the blended solvers. The solution with different refinement levels
and each Riemann solver is shown in Figure \ref{fig:rp_wet_bed}.
In Table \ref{table:rp_wet_bed}, we summarize the results of a convergence test.
Since the solution is non-smooth, we expect no more than first order convergence rates. Note that the
results with the entropy dissipative blended solver are comparable to the results using Roe's solver.
That is, imposing entropy dissipation via the blended Riemann solver does not degrade the high-order
accuracy properties of Roe's solver.
In contrast, the accuracy and convergence rates using Rusanov's solver are clearly degraded.

\begin{table}[!ht]\scriptsize
  \begin{center}
    \begin{tabular}{||c||c|c||c|c||c|c||} \hline
      & \multicolumn{2}{c||}{Roe's solver}
      &\multicolumn{2}{c||}{Rusanov's solver}
      &\multicolumn{2}{c||}{Blended solver} \\ \cline{2-7}
      $\Delta x$ & $E_1$ & rate & $E_1$ & rate & $E_1$ & rate \\ \hline
      1/50   & 4.22E-04 &  --  & 5.73E-04 &  --  & 4.24E-04 &  --  \\ 
      1/100  & 2.00E-04 & 1.07 & 3.35E-04 & 0.78 & 1.99E-04 & 1.08 \\
      1/200  & 1.11E-04 & 0.85 & 2.01E-04 & 0.74 & 1.10E-04 & 0.84 \\
      1/400  & 5.05E-05 & 1.13 & 1.18E-04 & 0.77 & 5.07E-05 & 1.12 \\
      1/800  & 2.60E-05 & 0.95 & 7.08E-05 & 0.74 & 2.61E-05 & 0.95 \\
      1/1600 & 1.29E-05 & 1.01 & 3.79E-05 & 0.90 & 1.29E-05 & 1.01 \\ \hline
    \end{tabular}
    \caption{Grid convergence study for the dam break problem over a wet bed
      using method \eqref{second-order_via_fluct} with different Riemann solvers.\label{table:rp_wet_bed}}
  \end{center}
\end{table}

\subsection{Flow past a cylinder}\label{sec:bow_shock}
In this section we consider the formation of a bow shock when a
uniform flow encounters a cylindrical obstacle.  This problem has
been studied previously in the context of carbuncle formation in
several works for the Euler equations and in
\cite{kemm2014note,bader2014carbuncle} for the shallow water equations.
We present results for existing solvers as a form of verification,
along with results for the new blended solver.  The main question of
interest is the formation of carbuncles.  It is known, for instance,
that the Roe solver incorrectly generates a carbuncle at the center
of the bow shock.

We model only the flow on the upstream side of the cylinder, since
our interest is in the resolution of the bow shock.  We take the domain
$[0,40]\times[0,100]$ with a cylinder of radius 20 centered at $(40,50)$.
Reflecting boundary conditions are imposed at the surface of the cylinder,
along with outflow conditions along the rest of the right edge of the domain.
The depth and velocity are set initially and (for all time) at the other
boundaries to $h_0=1$, $u_0=5$, for a Froude number of 5.  We use a $160 \times 400$
uniform Cartesian grid.
In Figure \ref{fig:steady_outflow}, we show results at $t=80$, after the flow has reached
a steady state.  With Roe's solver, negative values (of the water height) are generated at an early
time, leading to failure of the solver.  Therefore, we show results only for
Rusanov's, HLLEMCC by \cite{kemm2014note}, and the blended solvers.  These three methods give very similar
results.

\begin{figure}[!h]
  {\scriptsize
    \subfloat[Rusanov's solver.]{\includegraphics[scale=0.4]{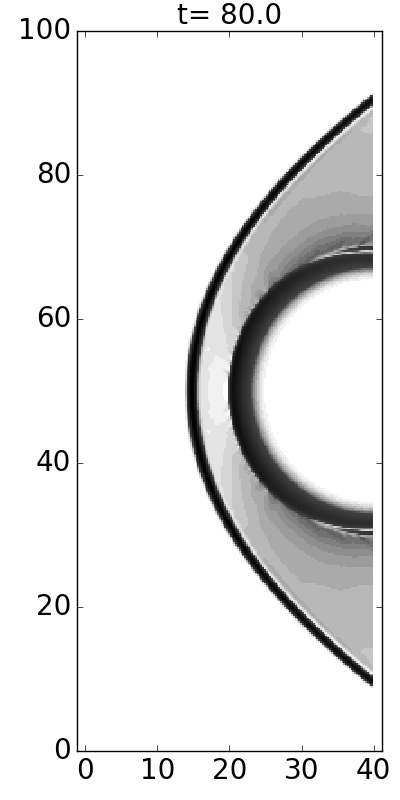}}
    \quad \quad \quad \quad \quad \quad
    \subfloat[HLLEMCC solver.]{\includegraphics[scale=0.4]{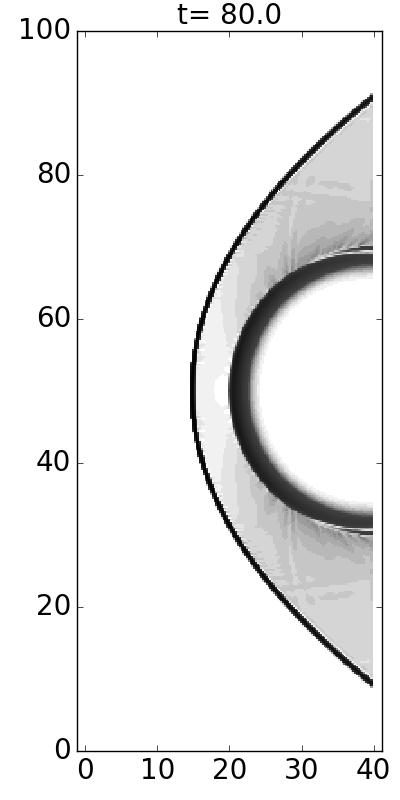}}
    \quad \quad \quad \quad \quad \quad
    \subfloat[Blended solver.]{\includegraphics[scale=0.4]{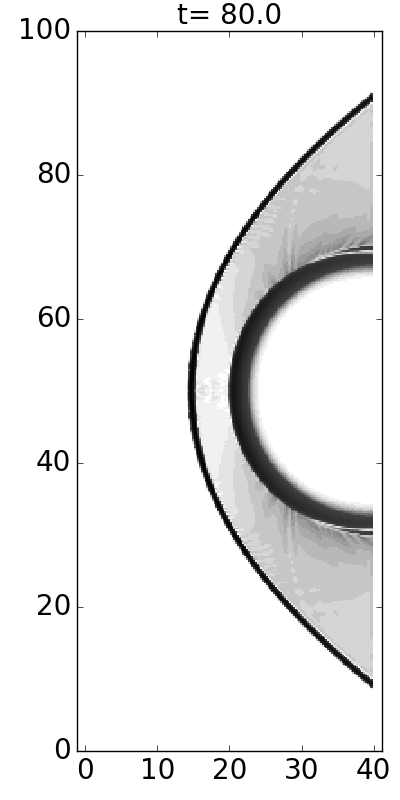}}
  }
  \caption{
    Flow past a cylinder.  Results for Rusanov's, HLLEMCC, and the blended solvers.
    \label{fig:cylinder_flow}}
\end{figure}

We have also conducted a more challenging version of this test, in which
the velocity within the domain is initially zero.  In that case, Rusanov's 
and the blended solvers give carbuncle-free results similar to Figure \ref{fig:steady_outflow},
but HLLEMCC exhibits a carbuncle.

\section{Numerical study of the circular hydraulic jump}\label{sec:chj}

\subsection{Semi-analytical steady solution under rotational symmetry}\label{sec:steady_chj}
In this section, we consider the initial boundary value problem consisting of the
shallow water model \eqref{eq:sw} in an annular domain
\begin{align} \label{eq:annulus}
r_\textup{jet} \le r \le r_\infty
\end{align}
where $r = \sqrt{x^2+y^2}$, with prescribed inflow at $r=r_\textup{jet}$
and prescribed outflow at $r=r_\infty$.
The domain and boundary conditions are rotationally symmetric.
By assuming rotational symmetry in \eqref{eq:sw}, one obtains the system
\begin{subequations} \label{eq:rsw}
\begin{align}
    (rh)_t + (rhu)_r &= 0, \label{mass1} \\
    (rhu)_t + (rhu^2)_r + r \left(\frac{1}{2}gh^2\right)_r &= 0, \label{mom1}
\end{align}
\end{subequations}
where the depth $h$ and radial velocity $u$ are functions of $r$ and $t$.
By direct integration one finds that steady-state solutions of
\eqref{eq:rsw} satisfy
\begin{subequations}\label{steady}
\begin{align}
    rhu & = C \\
    h'(r) & = \frac{h(r)}{\frac{g}{\beta^2} r^3 (h(r))^3 -r} = \frac{h(r)}{r} \cdot \frac{(F(r))^2}{1-(F(r))^2} \label{eq:dh}
\end{align}
\end{subequations}
for some $C$ independent of $r$.  Here $F(r)=|u(r)|/\sqrt{gh(r)}$ is the Froude number.
We see that two types of steady profiles exist, depending on whether the flow
is subcritical ($|F|<1$) or supercritical ($|F|>1$).  No smooth steady solution can
include both regimes, since the right hand side of \eqref{eq:dh} blows up when $F=1$.

\subsection{Numerical test: steady outflow}\label{sec:steady_outflow}
\begin{figure}[!h]
  \centering 
  \includegraphics[scale=0.15]{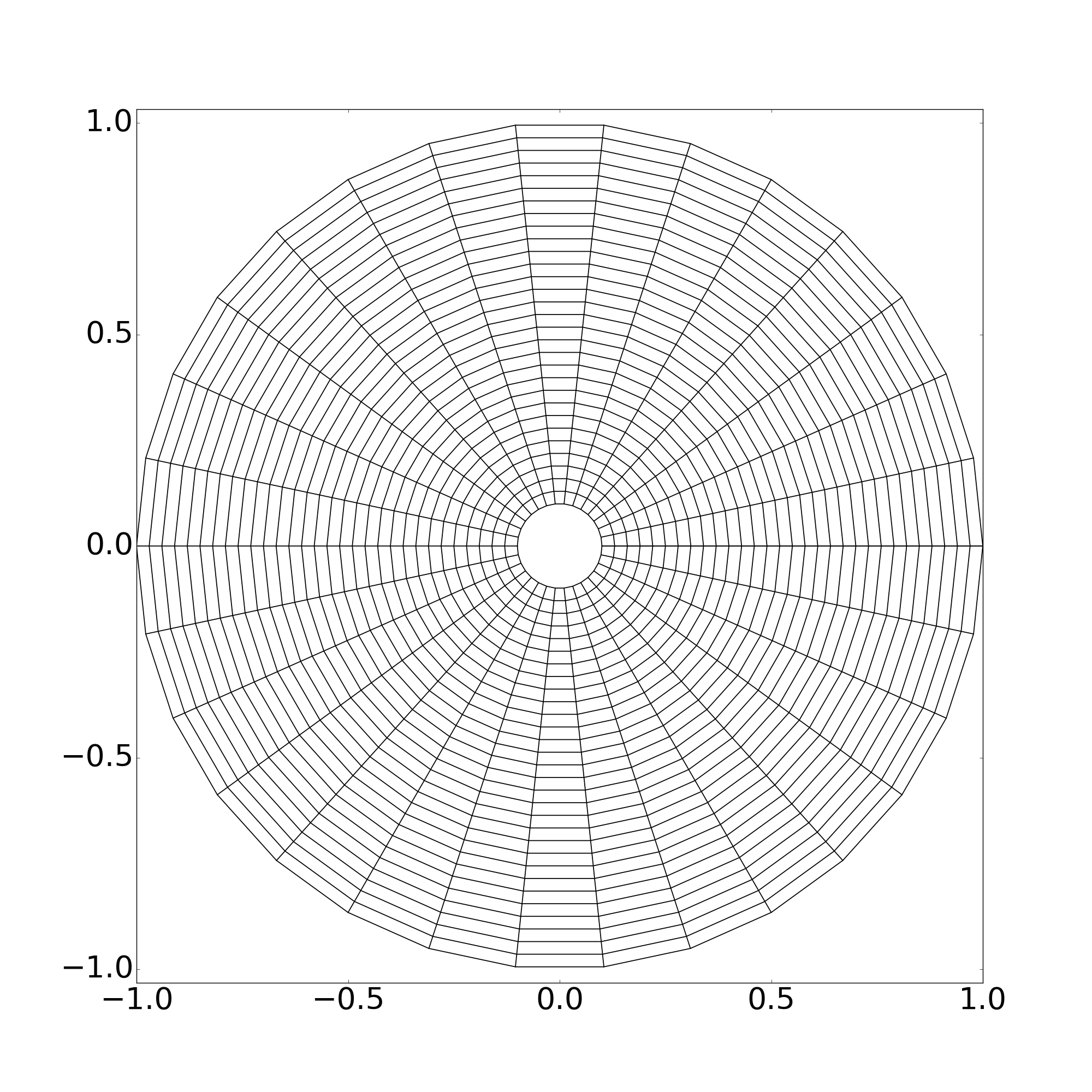}
  \caption{
    Example of a grid for the problem with the two-dimensional CHJ.
    \label{fig:mesh_chj}}
\end{figure}

\begin{figure}[!h]
  {\scriptsize
    \includegraphics[scale=0.29]{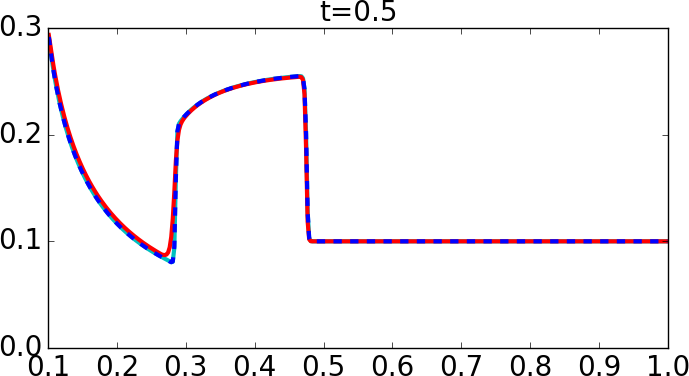}
    \qquad
    \includegraphics[scale=0.29]{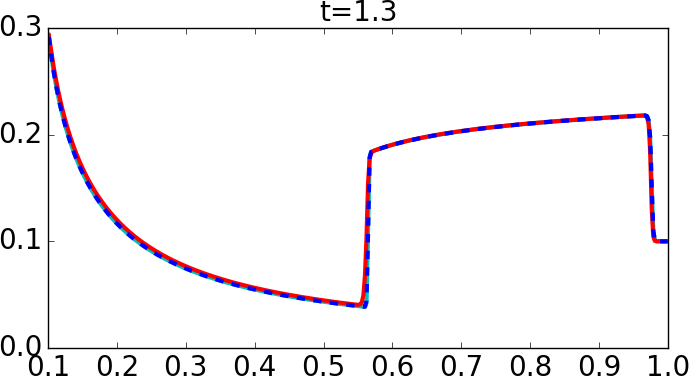}
    \qquad
    \includegraphics[scale=0.29]{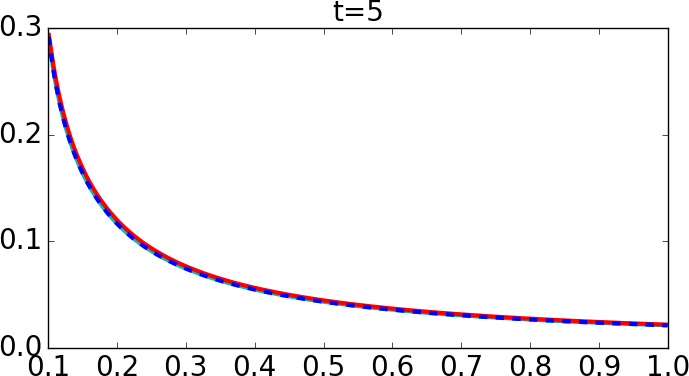}

    \vspace{15pt}
    \includegraphics[scale=0.29]{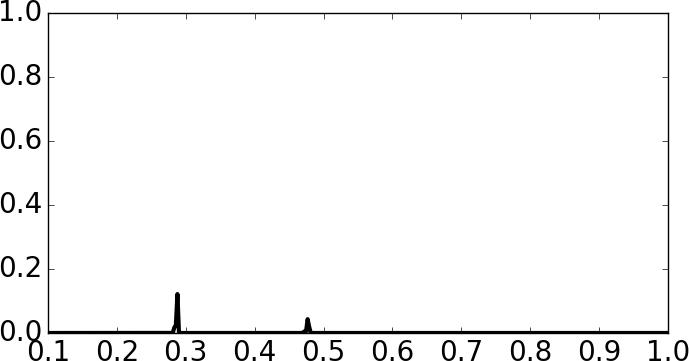}
    \qquad
    \includegraphics[scale=0.29]{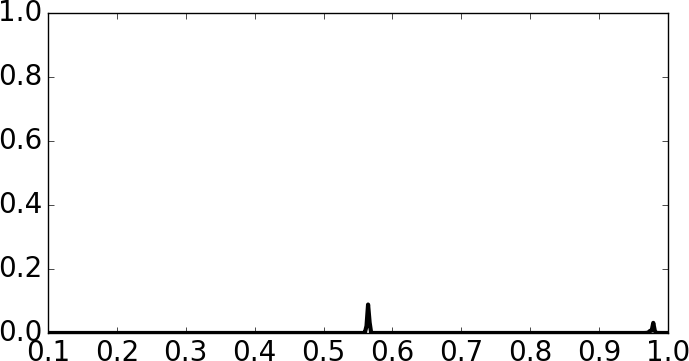}
    \qquad
    \includegraphics[scale=0.29]{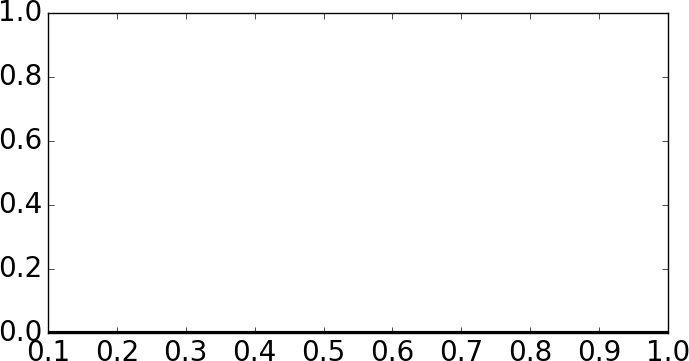}
  }
  \caption{
    Steady outflow problem using method \eqref{second-order_via_fluct} with
    (cyan) Roe's, (red) Rusanov's and (dashed blue) the blended Riemann solvers.
    In the second row we plot $\theta_i$, which is given by \eqref{Ri}.
    In all simulations we take $\Delta x=1/400$.
    \label{fig:steady_outflow}}
\end{figure}

We now test the numerical methods by using a time-dependent simulation
to compute the steady flow solution just described, 
in the annulus $r\in(0.1,1)$ with constant inflow at $r=0.1$ and outflow at $r=1$.
The initial condition is
$h(r,t=0)=0.1$, $hu(r,t=0)=0$, the inner boundary condition is $h(0.1,t)=0.3$, $hu(0.1,t)=0.75$,
and the outer boundary condition is set to outflow
(see \cite[\S 21.8.5]{leveque2002finite} for details).
The computational mesh is logically quadrilateral, of the type shown in
Figure \ref{fig:mesh_chj}.
Regardless of the initial condition, the exact solution converges to a
steady state profile given by one of the two solutions of \eqref{steady},
corresponding to subcritical or supercritical flow.  In the present case we
have imposed a supercritical inflow.
In Figure \ref{fig:steady_outflow}, we show the solution and $\theta_i$ at different times using
the second-order method \eqref{second-order_via_fluct}
with Roe's, Rusanov's and the blended Riemann solvers.  
Additionally, in Table \ref{table:steady_outflow}, we summarize the results of a convergence study
based on methods \eqref{first-order_via_fluct} and
\eqref{second-order_via_fluct}, using the same Riemann solvers.
Although the chosen initial
condition leads initially to shock formation, the steady state is smooth and close to
second order convergence is observed for the Roe and blended solvers.

\begin{table}[!ht]\scriptsize
  \begin{center}
    \begin{tabular}{||c||c|c||c|c||c|c||c|c||c|c||c|c||} \hline
      & \multicolumn{6}{c||}{First-order method \eqref{first-order_via_fluct}}
      & \multicolumn{6}{c||}{Second-order method \eqref{second-order_via_fluct}} \\ \cline{2-13}
      & \multicolumn{2}{c||}{Roe's solver}
      &\multicolumn{2}{c||}{Rusanov's solver}
      &\multicolumn{2}{c||}{Blended solver}
      & \multicolumn{2}{c||}{Roe's solver}
      &\multicolumn{2}{c||}{Rusanov's solver}
      &\multicolumn{2}{c||}{Blended solver}
      \\ \cline{2-13}
      $\Delta x$ & $E_1$ & rate & $E_1$ & rate & $E_1$ & rate & $E_1$ & rate & $E_1$ & rate & $E_1$ & rate\\ \hline
      1/50   & 2.16E-4 &  --  & 9.97E-4 &   -- & 2.16E-4 & --   & 4.13E-5 &   -- & 7.32E-4 &  --  & 4.15E-5 & --   \\
      1/100  & 1.14E-4 & 0.92 & 6.38E-4 & 0.64 & 1.14E-4 & 0.92 & 1.47E-5 & 1.48 & 4.43E-4 & 0.73 & 1.48E-5 & 1.48 \\
      1/200  & 5.86E-5 & 0.95 & 3.88E-4 & 0.72 & 5.87E-5 & 0.95 & 4.81E-6 & 1.61 & 2.57E-4 & 0.79 & 4.83E-6 & 1.61 \\
      1/400  & 2.98E-5 & 0.97 & 2.26E-4 & 0.78 & 2.98E-5 & 0.97 & 1.42E-6 & 1.76 & 1.44E-4 & 0.83 & 1.42E-6 & 1.76 \\
      1/800  & 1.50E-5 & 0.98 & 1.27E-4 & 0.83 & 1.50E-5 & 0.98 & 4.16E-7 & 1.76 & 7.81E-5 & 0.88 & 4.18E-7 & 1.76 \\
      1/1600 & 7.53E-6 & 0.99 & 6.90E-5 & 0.88 & 7.53E-6 & 0.99 & 1.14E-7 & 1.86 & 4.12E-5 & 0.92 & 1.15E-7 & 1.86 \\ \hline
    \end{tabular}
    \caption{Grid convergence study for the steady outflow problem
      using methods \eqref{first-order_via_fluct} and \eqref{second-order_via_fluct}
      with different Riemann solvers.\label{table:steady_outflow}}
  \end{center}
\end{table}

\subsection{Location of the jump} \label{sec:jumploc}
The steady, rotationally-symmetric circular hydraulic jump involves supercritical
flow for $r<r_0$ and subcritical flow for $r>r_0$, where $r_0$ is the jump radius.
The jump itself takes the form of a stationary shock wave.  The Rankine-Hugoniot jump
conditions specify that for such a shock,
\begin{align} \label{eq:RH}
    h_+ - h_- & = \frac{-3h_- + \sqrt{h_-^2 + 8 h_- u_-^2/g}}{2} = \frac{3h_-}{2}\left(\sqrt{1+\frac{8}{9}(F_-^2-1)}-1\right),
\end{align}
where the subscripts $+, -$ denote states just inside or outside the jump radius, respectively.

A steady-state, rotationally symmetric solution can be given for an annular region with prescribed
flow at the inner and outer boundaries as follows:
\begin{enumerate}
    \item Specify the depth and velocity at the inner boundary (near the jet) and outer boundary.
    \item Integrate \eqref{eq:dh} from both boundaries.
    \item Find a radius $r_0$ where the matching condition \eqref{eq:RH} is satisfied.
\end{enumerate}
Due to the nature of solutions of \eqref{eq:dh}, it can be shown that the required jump
radius $r_0$ always exists if the prescribed flow is supercritical at the inner boundary
and subcritical at the outer boundary.

\begin{figure}[!h]
  \centering
  \subfloat[$F(r_\jet)\approx 1.37$\label{fig:ic_regime_I}]{\includegraphics[scale=0.3]{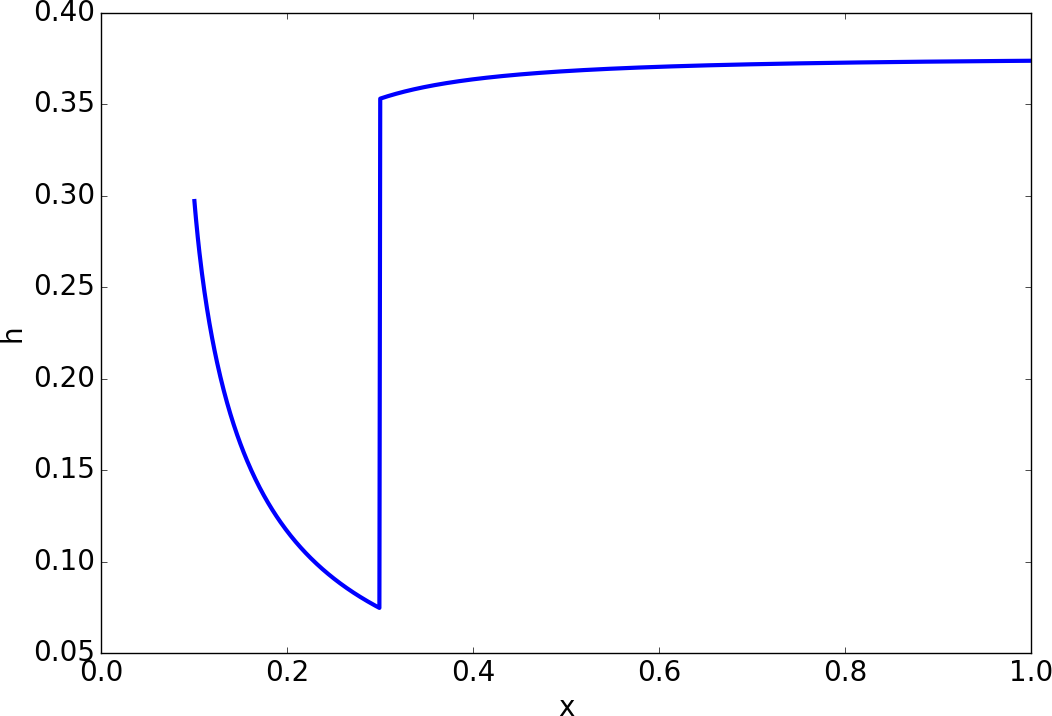}} \qquad
  \subfloat[$F(r_\jet)\approx 27.39$\label{fig:ic_regime_II}]{\includegraphics[scale=0.3]{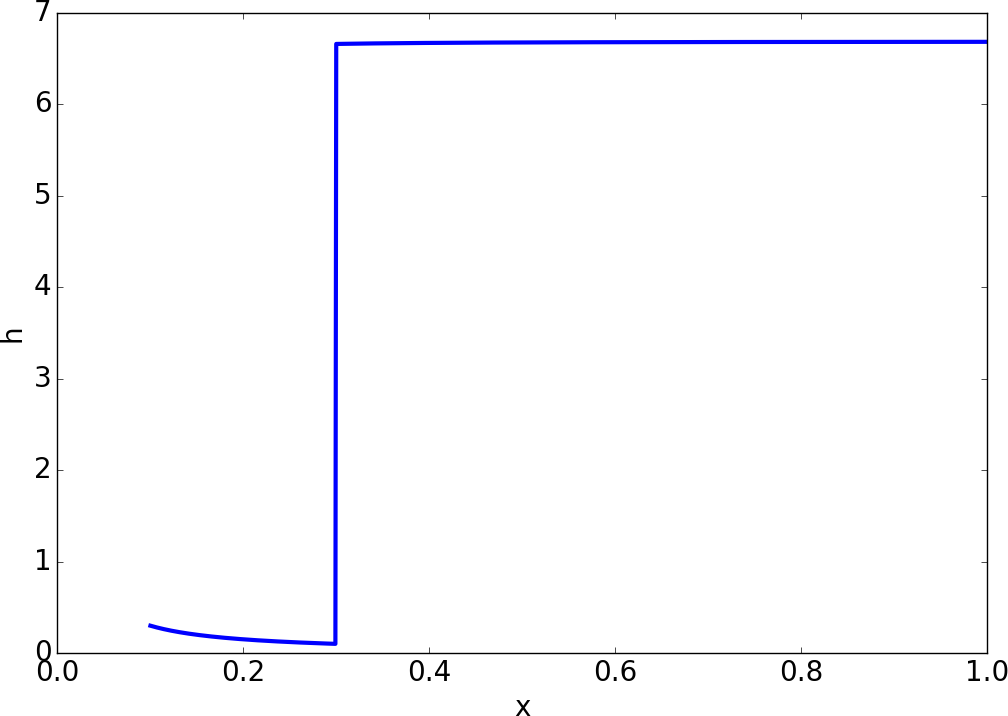}}
  \caption{Radially symmetric CHJs created by solving \eqref{steady} with a jump given by \eqref{eq:RH}.
    We show a slice along $y=0$.
    In Section \ref{sec:regime_i}, we consider the CHJ in (a) as initial condition;
    and in Sections \ref{sec:regime_ii} and \ref{sec:regime_iiwp},
    we consider the CHJ in (b) as initial condition. 
    \label{fig:regimes}}
\end{figure}

In this section we described how the location of the jump for a steady, rotationally-symmetric CHJ
is determined by the boundary conditions. Following similar steps, one can choose inner boundary conditions
and find outflow boundary conditions that lead to a CHJ at a prescribed location.
This a convenient approach to construct initial conditions for numerical experiments at different flow regimes. 
Let us consider two flow regimes and construct the corresponding CHJs, which we use in the following sections.
Consider the following boundary conditions:
\begin{subequations}\label{bcs_chj}
  \begin{align}
  h(x,y,t)=h_\jet, \quad
  u(x,y,t)=|\bfu_\jet| \left(\frac{x}{r_\jet}\right), \quad
  v(x,y,t)=|\bfu_\jet| \left(\frac{y}{r_\jet}\right), \qquad \sqrt{x^2+y^2}=r_\jet \\
  h(x,y,t)=h_\out, \quad 
  u(x,y,t)= \frac{\beta }{r_\out h}\left(\frac{x}{r_\out}\right), \quad
  v(x,y,t)= \frac{\beta }{r_\out h}\left(\frac{y}{r_\out}\right), \qquad \sqrt{x^2+y^2}=r_\out,
  \end{align}
\end{subequations}
where $h_\jet=0.3$, $r_\jet=0.1$, $r_\out=1$ and $\beta=r_\jet h_\jet|\bfu_\jet|$.
We choose $|\bfu_\jet|$ and $h_\out$ such that the steady roationally-symmetric
solution involves a symmetric shock at $r_s=0.3$; see Table \ref{table:regimes}.
In Figure \ref{fig:regimes}, we show the water depth $h$ along $y=0$ for these regimes.

\begin{table}[!ht]
  \begin{center}
    \begin{tabular}{||c||c|c|c||} \hline
      & $F(r_\jet)$ &  $|\bfu_\jet|$ & $h_\out$ \\ \hline
      Regime I & 1.37 & 0.75 & 0.37387387318873766 \\ \hline
      Regime II & 27.39 & 15 & 6.6845019298155357 \\ \hline
    \end{tabular}
    \caption{Boundary data $|u_\jet|$ and $h_\out$ which along with \eqref{bcs_chj}
      produce CHJs located at $r_s=0.3$ for two different flow regimes. \label{table:regimes}}
  \end{center}
\end{table}

Numerical tests suggest that the solution of \eqref{eq:rsw} rapidly approaches that
just described under general initial conditions as long as the inflow at the jet is
supercritical and the outflow at $r_\infty$ is subcritical.  Subcritical outflow
can be enforced by an appropriate outer boundary condition.  This solution remains steady
if the rate of inflow and outflow are matched.  The stability of this solution in
the face of non-rotationally-symmetric perturbations is an important question not
only in the shallow water context but for more realistic fluid models and physically.
It will play an important role in the results we present below.
The rotationally-symmetric steady state is a useful initial condition for studies
of CHJ stability.

\subsection{The circular hydraulic jump in two dimensions}\label{sec:2D_chj}
Let us finally consider the numerical experiments for the CHJ in two dimensions.
The domain is again given by the annulus \eqref{eq:annulus}, but now we solve
the fully 2D shallow water equations \eqref{eq:sw}.  For all of the following
experiments, we use a mesh with 1000 cells in each (radial and angular) direction.
The boundary conditions at the jet and the outer boundary are given by \eqref{bcs_chj}.
By adjusting the boundary conditions $|\bfu_\jet|$ and $h_\out$ we can study different flow regimes.
We focus on the two cases in Table \ref{table:regimes}.
%
%
For most of the experiments we show a Schlieren plot for the water height.
That is, we plot $||\nabla h||_{\ell^2}$ with a greyscale logarithmic colormap.

\subsubsection{Regime I ($F_\jet\approx1.37$)}\label{sec:regime_i}
\begin{figure}[!h]
  \centering
  \subfloat[Solution at (left) $t=1$, (middle) $t=1.5$ and (right) $t=3$ using Roe's Riemann solver.
    The main plots are Schlieren plots of the depth $h$, while the inset in the right
    figure shows the momentum in the radial direction.\label{fig:chj_r1_roe}]{
    \begin{tabular}{ccc} 
      \includegraphics[scale=0.3]{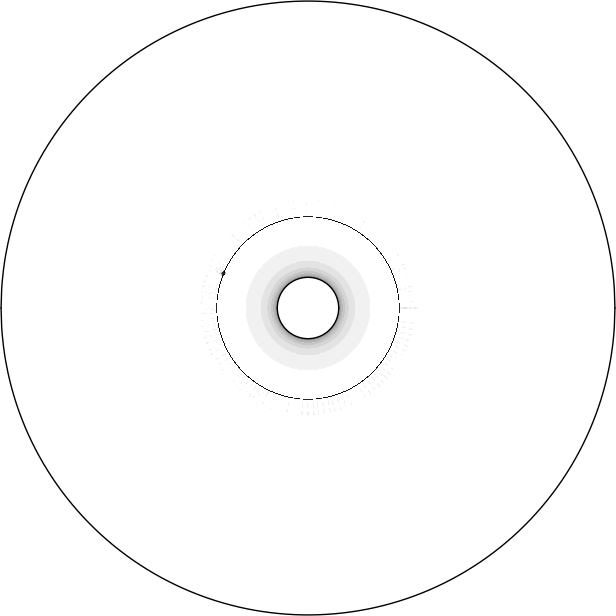} &
      \includegraphics[scale=0.3]{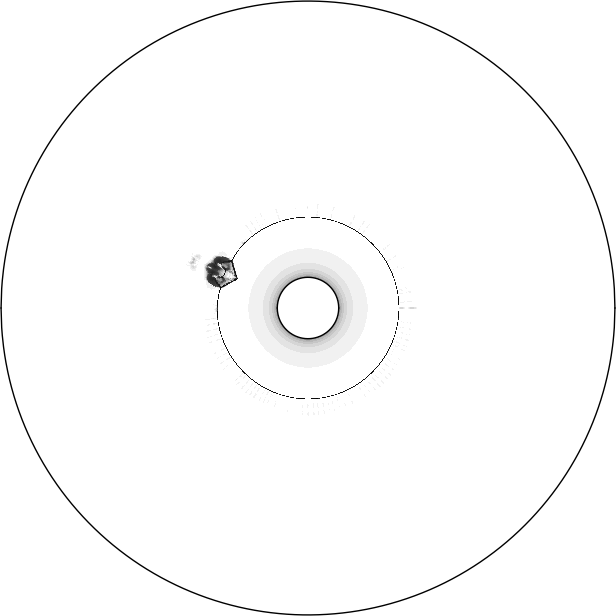} &
      \includegraphics[scale=0.3]{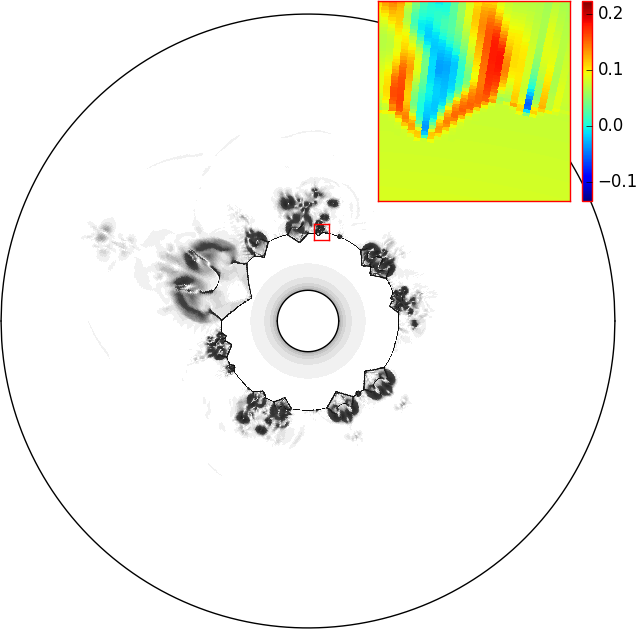}
    \end{tabular}
  }
  
  \subfloat[Solution at $t=3$ using (left) Rusanov's,
    (upper-right) HLLEMCC and (bottom-right) the blended solvers.\label{fig:chj_r1_diff}]{
    \includegraphics[scale=0.3]{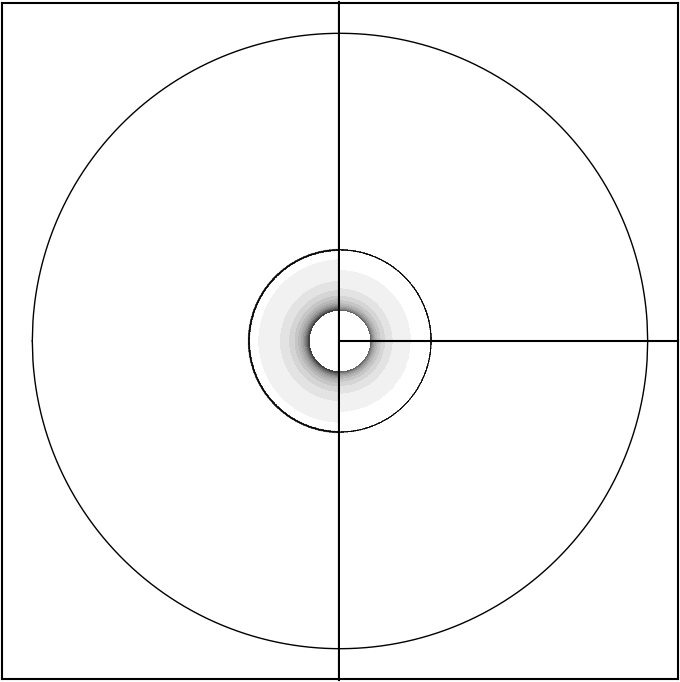}
  }
  \qquad\qquad
  \subfloat[Solution at $t=3$ along $y=0$ using the same solvers as in (b).]{
    \includegraphics[scale=0.3]{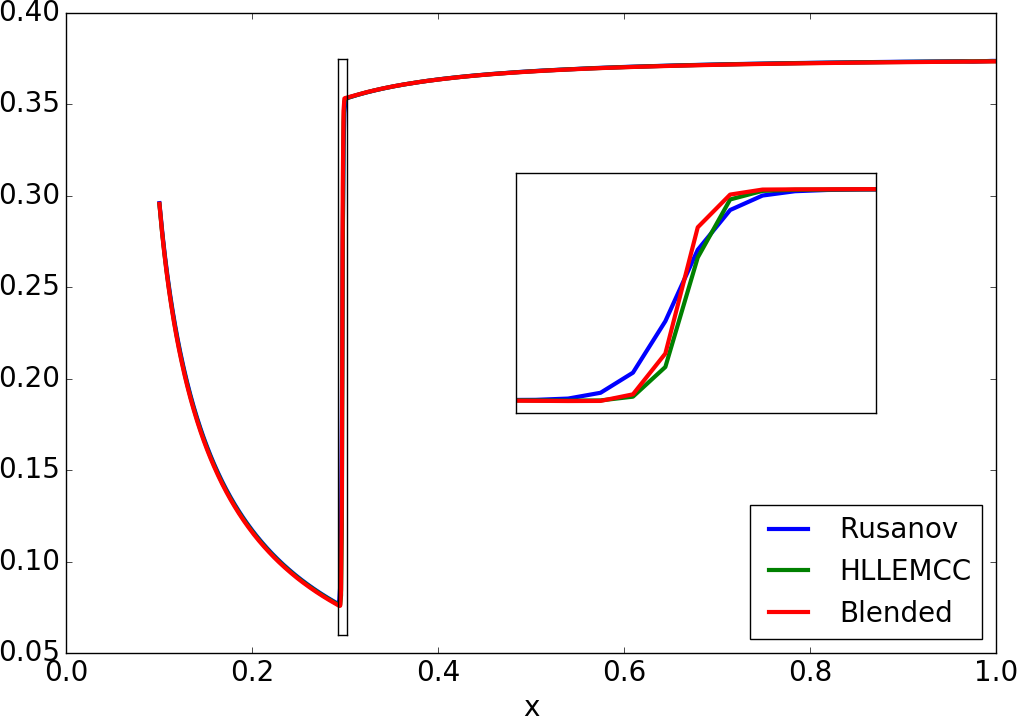}}
  
    \caption{Simulation of a CHJ with boundary conditions given by \eqref{bcs_chj} and \eqref{bcs_chj_r1}.    
      We consider method \eqref{second-order_via_fluct} with different Riemann solvers.
      In (a) and (b) we show the Schlieren plot for the water height $h$.
    \label{fig:chj_r1}}
\end{figure}

In this case, the boundary conditions are given by \eqref{bcs_chj} with
\begin{align}\label{bcs_chj_r1}
  |\bfu_\jet|=0.75, \qquad h_\out=0.37387387318873766. 
\end{align}
The initial condition is a circular hydraulic jump located at $r_s=0.3$; see Figure \ref{fig:ic_regime_I}. 
In Figure \ref{fig:chj_r1_roe}, we show the solution at different times using
method \eqref{second-order_via_fluct} with Roe's Riemann solver.
The solution clearly develops carbuncle instabilities, which are evident in the inset figure.
In Figure \ref{fig:chj_r1_diff}, we show solutions at $t=3$ for
Rusanov's, HLLEMCC, and the blended Riemann solvers.
For this test case, these three solvers give qualitatively similar solutions,
all of which are free from carbuncles.

\subsubsection{Regime II ($F_\jet\approx27.39$)}\label{sec:regime_ii}
\begin{figure}[!h]
  \centering
  \subfloat[Solution at (left) $t=0.09$, (middle) $t=0.1$ and (right) $t=0.11$ using Roe's Riemann solver.\label{fig:chj_r2_roe}]{
    \begin{tabular}{ccc}
      \includegraphics[scale=0.3]{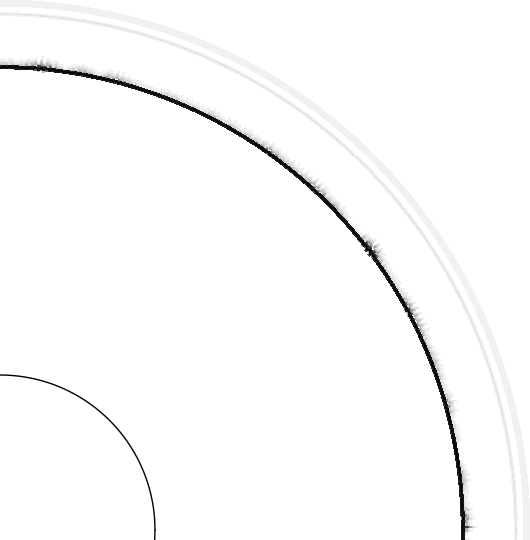} &
      \quad\includegraphics[scale=0.3]{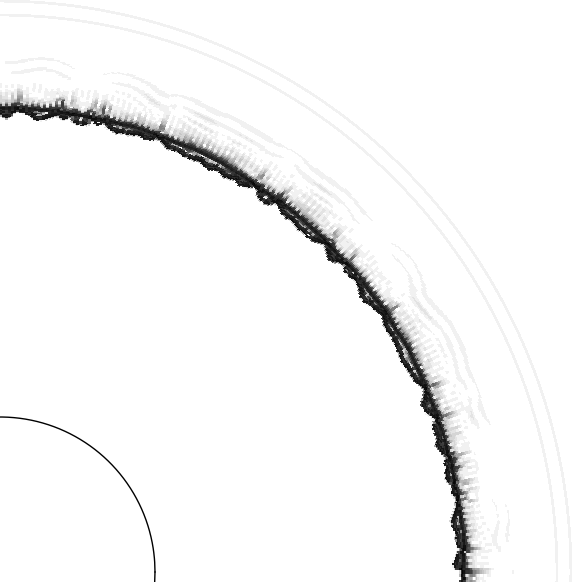} \quad &
      \includegraphics[scale=0.3]{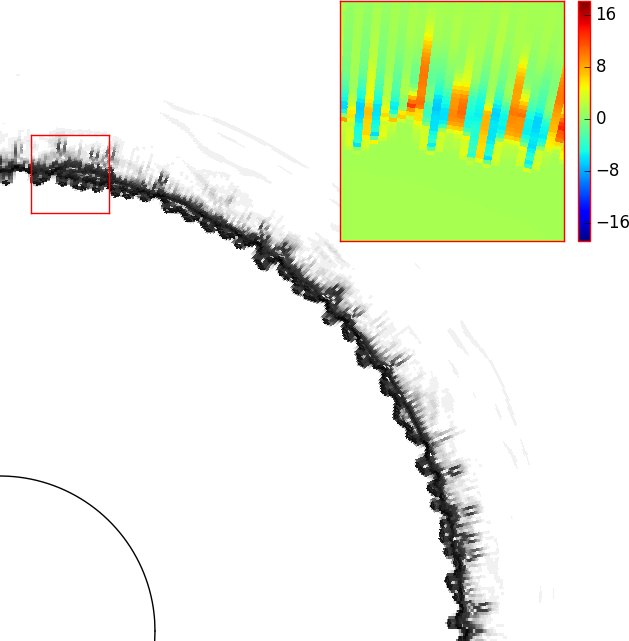}
    \end{tabular}
  }

  \vspace{10pt}
  \subfloat[Solution at $t=0.11$ using (left) Rusanov's and (right) the blended solvers.\label{fig:chj_r2_diff}]{
    \begin{tabular}{ccc}
      \includegraphics[scale=0.3]{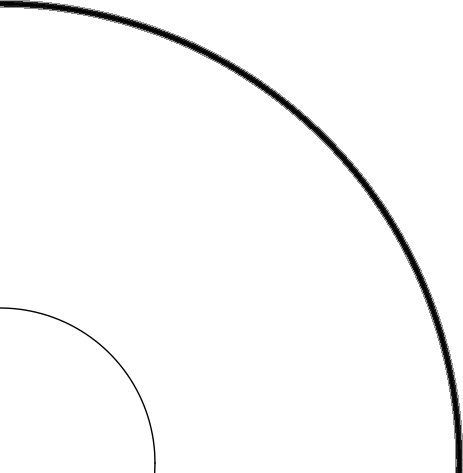} \qquad &
      \qquad \includegraphics[scale=0.3]{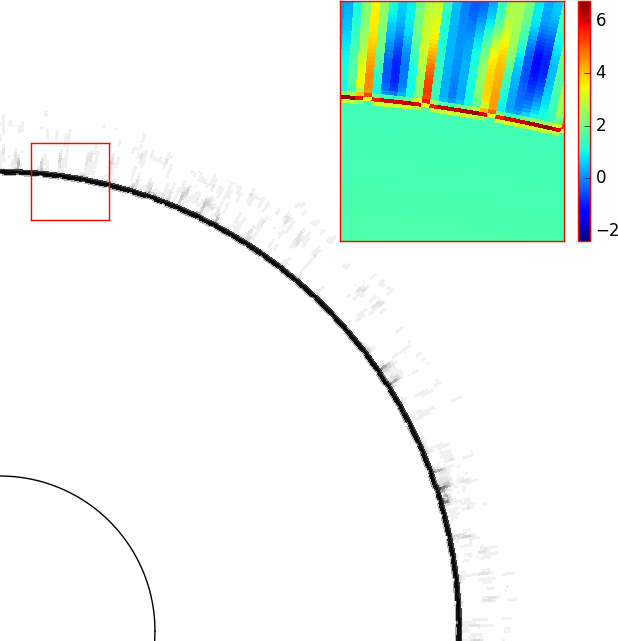}
    \end{tabular}
    }
  \caption{CHJ with boundary conditions given by \eqref{bcs_chj} and \eqref{bcs_chj_r2}. 
    We consider method \eqref{second-order_via_fluct} with different Riemann solvers.
    In all cases we show the Schlieren plot for the water height $h$.
    The inset figures show the momentum in the radial direction.}
\end{figure}

We now consider a higher-Froude number regime. The boundary conditions are
given by \eqref{bcs_chj} with 
\begin{align}\label{bcs_chj_r2}
  |\bfu_\jet|=15, \qquad h_\out=6.6845019298155357.
\end{align}
The initial condition is a circular hydraulic jump located at $r_s=0.3$; see Figure \ref{fig:ic_regime_II}.
In Figure \ref{fig:chj_r2_roe}, we show the solution at different times using method
\eqref{second-order_via_fluct} with Roe's Riemann solver. Again the solution develops
carbuncle instabilities, which are clearly seen in the inset figure at $t=3$.
For this regime, we obtained negative values for the water height with
the HLLEMCC solver, which lead to failure of the solver. Therefore, in Figure
\ref{fig:chj_r2_diff} we only show results with Rusanov's and the blended
solvers.  

\begin{figure}[!h]
  \centering
    \includegraphics[scale=0.3]{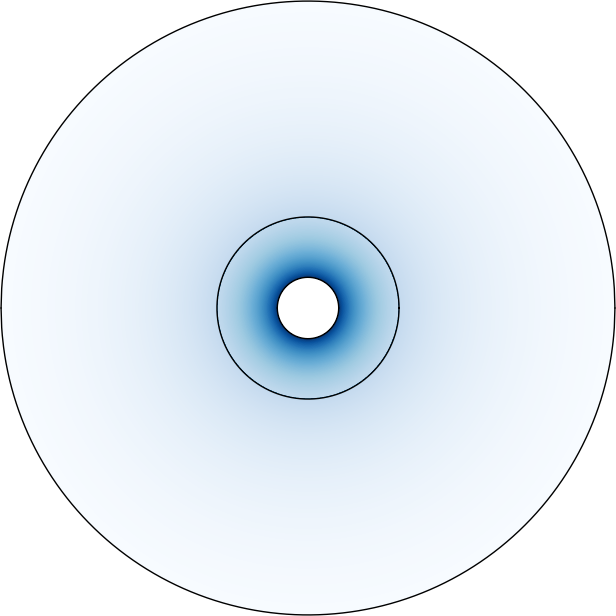} \quad
    \quad \includegraphics[scale=0.3]{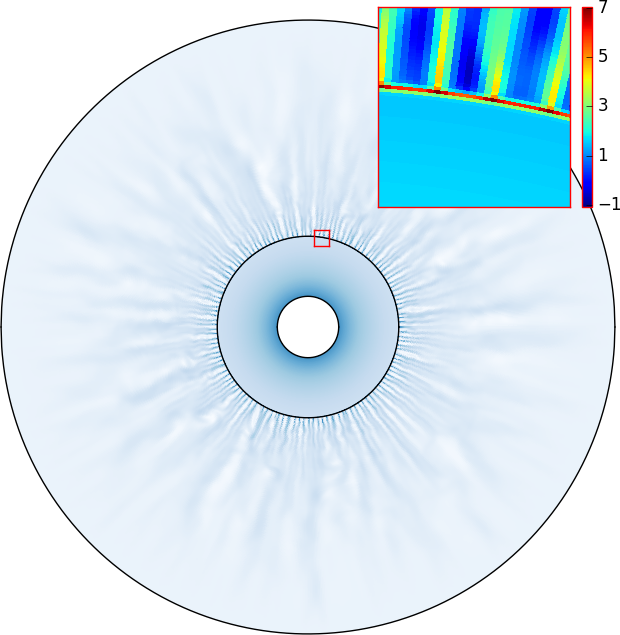}
  \caption{CHJ at $t=5$ with boundary conditions given by \eqref{bcs_chj} and \eqref{bcs_chj_r2}. 
    We consider method \eqref{second-order_via_fluct} with (left) Rusanov's
    and (right) the blended solvers. 
    In all cases we superimpose the Schlieren plot for the water height $h$
    an the magnitude of the momentum with a uniform (white-to-blue) colormap. 
    The scale for colormap is (white) $0.45$ to (blue) $4.5$, 
    which corresponds to the minimum and maximum values at $t=0$.
    The inset figure shows the momentum in the radial direction.
    \label{fig:chj_r2_later}}
\end{figure}

The solutions obtained with these solvers show no carbuncles.  The Rusanov
solution remains very close to the initial symmetric equilibrium state,
whereas the blended solver solution includes perturbations that appear
just downstream from the jump.
In Figure \ref{fig:chj_r2_later}, we show the Rusanov and blended solutions at
a much later time of $t=5$.  In addition to the Schlieren plot of the water
height, we superimpose a color plot of the magnitude of the momentum. 
It is evident that the visible perturbations in the blended solution are
completely dissipated in the Rusanov solution, at least when using the grid employed here.

It is natural to ask whether these perturbations are meaningful; i.e., whether
the symmetric equilibrium is unstable.  To investigate this, we conduct one
more test.

\subsubsection{Regime II with random perturbations}\label{sec:regime_iiwp}
\begin{figure}[!h]
  \centering
  \subfloat[Rusanov's Riemann solver\label{fig:chj_r2wp_rusanov}]{
    \begin{tabular}{cccc}
      $t=2$ & $t=3$ & $t=4$ & $t=5$ \\
      \includegraphics[scale=0.24]{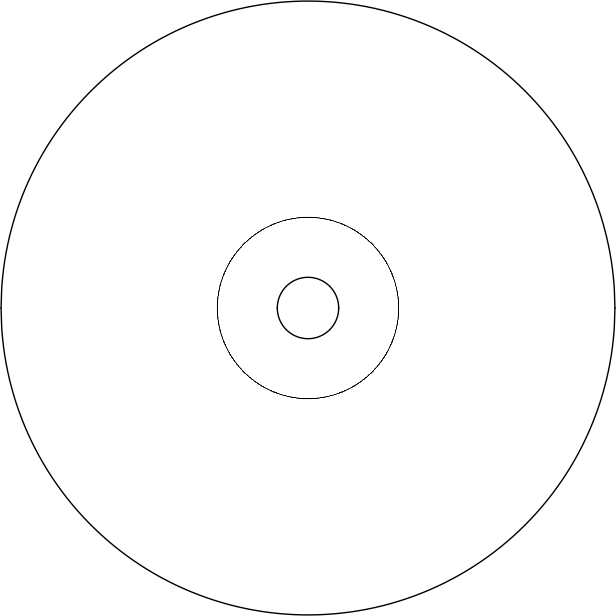} &
      \includegraphics[scale=0.24]{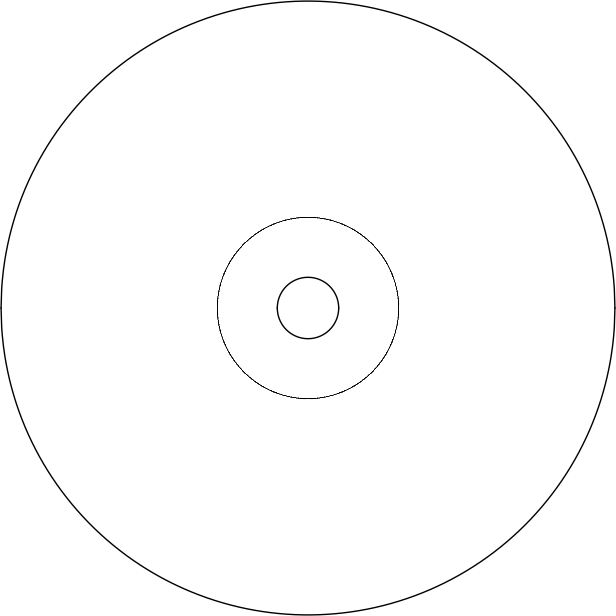} &
      \includegraphics[scale=0.24]{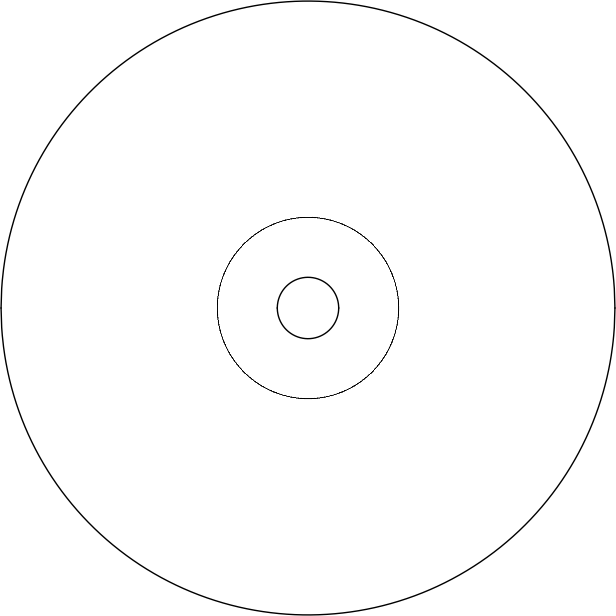} &
      \includegraphics[scale=0.24]{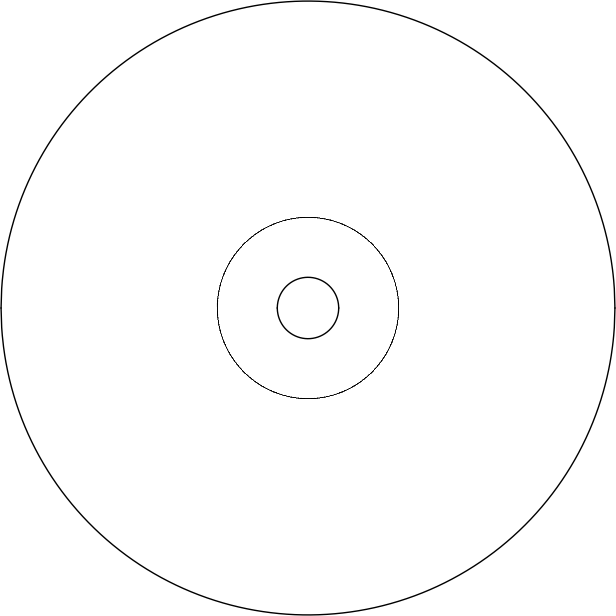}
    \end{tabular}
  }
  
  \subfloat[Blended Riemann solver\label{fig:chj_r2wp_blended}]{
    \begin{tabular}{cccc}
      \includegraphics[scale=0.24]{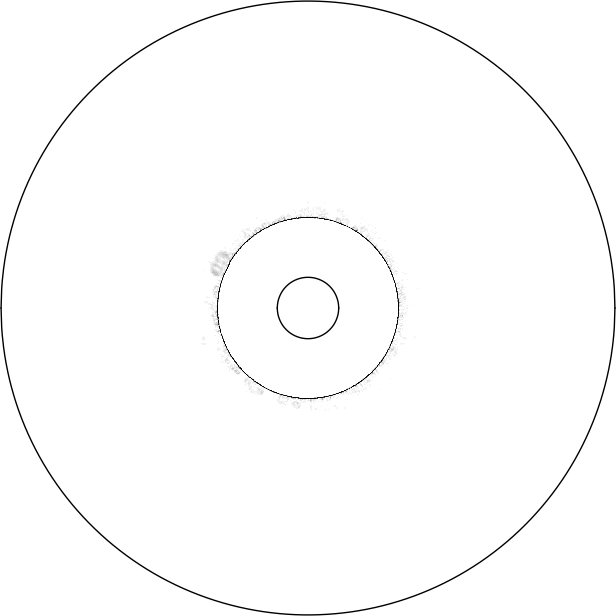} &
      \includegraphics[scale=0.24]{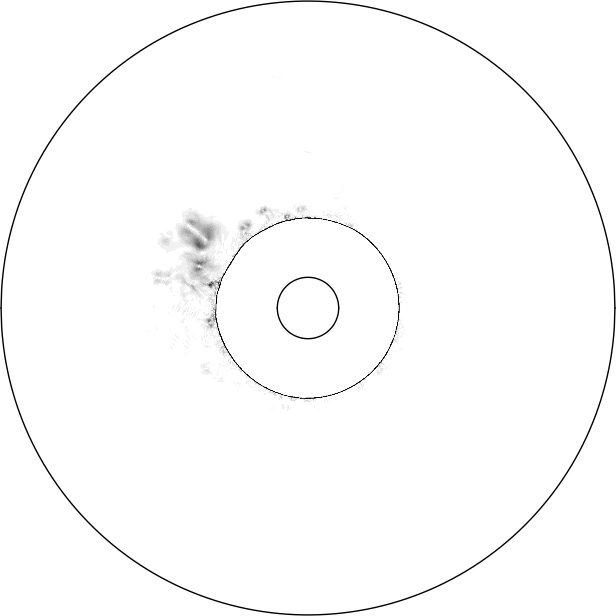} &
      \includegraphics[scale=0.24]{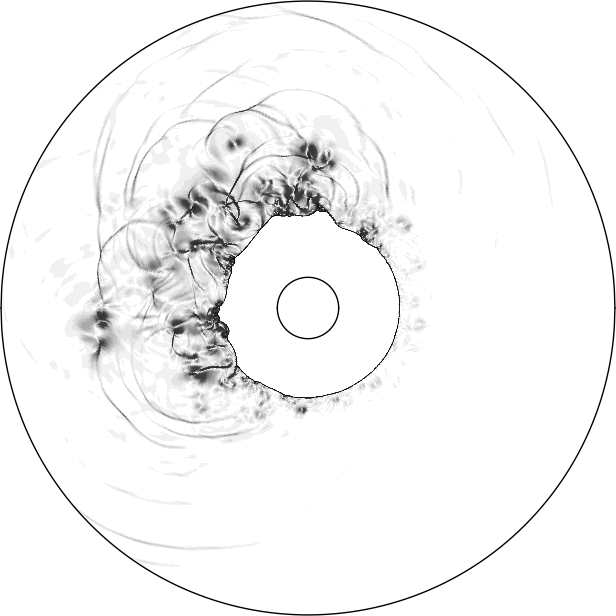} &
      \includegraphics[scale=0.24]{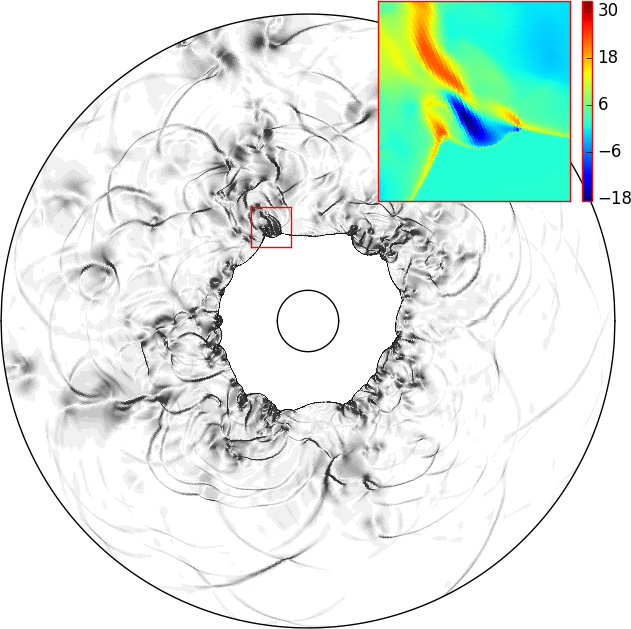}
    \end{tabular}
  }
  \caption{CHJ at different times with a random perturbation at the inflow boundary.
    We consider method \eqref{second-order_via_fluct} with different Riemann solvers. 
    In all cases we show the Schlieren plot for the water height $h$.
    The inset figure shows the momentum in the radial direction.}
\end{figure}

Given the symmetry of the grid and initial data in the section above,
the non-rotationally-symmetric perturbations seen when using the blended
solver above must arise due to the influence of numerical errors.
In order to understand whether these are evidence of a true instability,
we now introduce a non-symmetric perturbation at the inflow boundary.
We use the same mean values \eqref{bcs_chj_r2}, but now we set
\begin{subequations}
  \begin{align}
    h(x,y,t)=\tilde h_\jet, \quad
    u(x,y,t)=|\tilde \bfu_\jet| \left(\frac{x}{r_\jet}\right), \quad
    v(x,y,t)=|\tilde \bfu_\jet| \left(\frac{y}{r_\jet}\right), \qquad r=r_\jet,
  \end{align}
\end{subequations}
with $\tilde h_\jet = \frac{h_\jet}{1+\epsilon(x,y,t)}$ and $|\tilde\bfu_\jet|=|\bfu_\jet|(1+\epsilon(x,y,t))$.
Here $\epsilon(x,y,t)$ is chosen at each inflow boundary ghost cell and at each
time step as an i.i.d. random variable from the uniform distribution $[-0.01,0.01]$.
In this case, due to the large Froude number, the random perturbations in the inflow
may trigger physical instabilities, which we want to preserve. At the same time, we
do not want the solution to develop carbuncle instabilities.
In figures \ref{fig:chj_r2wp_rusanov} and \ref{fig:chj_r2wp_blended}, we show the solution
at different times considering method \eqref{second-order_via_fluct} with Rusanov's and
the blended Riemann solvers, respectively.
In the right panel of Figure \ref{fig:chj_r2wp_blended}, we show an inset figure 
with the momentum in the radial direction. Note that even though the solution is highly unstable 
(with the blended Riemann solver), no visible carbuncle instabilities are developed.

\section{Conclusions}\label{sec:conclusion}
In this work we have introduced a new Riemann solver for the shallow water
equations, described in Section \ref{sec:blended_rs}.  Through numerical tests we have
shown that the solver gives accuracy similar to that of Roe's method and
robustness similar to that of Rusanov's method.  Although the full
second-order method we have proposed is not provably entropy stable or
positivity preserving, it gives improved results for test problems
where both of these properties are important.  The approach used in Section
\ref{sec:blended_rs} could be applied to enforce entropy stability for any Riemann
solver of the type used in Clawpack.  The same techniques could also
be used more generally to avoid carbuncles in the solution of the
Euler equations.

We have introduced two new test problems for numerical shallow water solvers,
both consisting of flow in an annulus, with inflow from a jet
at the center and outflow at the outer boundary.  The first test
problem, described in Section \ref{sec:steady_outflow}, has a smooth solution that can
be computed by solving an ODE and thus serves as a useful test of
accuracy.  The second problem, the circular hydraulic jump, involves
a standing shock wave that can be physically unstable but is
also susceptible to the numerical carbuncle instability.
Additionally, it involves high-velocity low-depth flow regions where
it is challenging to maintain positivity.
This makes it an excellent problem for testing that schemes are both
robust and not overly dissipative.

{\bf Acknowledgment.}  We thank Prof. Friedemann Kemm for sharing helpful code
with us and for reviewing an early version of this manuscript.
\bibliographystyle{plain}
\bibliography{refs}

\end{document}